\documentclass[a4paper,onecolumn,11pt,unpublished]{quantumarticle}
\pdfoutput=1

\usepackage[utf8]{inputenc}
\usepackage[english]{babel}
\usepackage[T1]{fontenc}
\usepackage[numbers,sort&compress]{natbib}

\usepackage{hyperref}
\usepackage{amsmath}
\usepackage{tikz}
\usepackage{lipsum}
\usepackage{graphicx}
\usepackage{subfig}
\usepackage{booktabs}
\usepackage{tabularx}
\usepackage{listings}
\usepackage{caption}

\newcommand{\Tesseract}{\textit{Tesseract}}
\newcommand{\QEC}{\textit{QEC}}
\newcommand{\astar}{{\textit{A*}}}
\newcommand{\Cpp}{\textit{C++}}

\newcommand{\getdetcost}{\texttt{get\_detcost}}
\newcommand{\vbool}{\texttt{std::vector<bool>}}
\newcommand{\vchar}{\texttt{std::vector<char>}}

\newcommand{\optone}{\textit{Optimization 1}}
\newcommand{\opttwo}{\textit{Optimization 2}}
\newcommand{\optthree}{\textit{Optimization 3}}
\newcommand{\optfour}{\textit{Optimization 4}}

\begin{document}
\title{Accelerating the Tesseract Decoder for Quantum Error Correction}

\author{Dragana Grbic}
\affiliation{Google Quantum AI, Venice, CA, 90291}
\affiliation{Department of Computer Science, Rice University, Houston, TX 77005}
\email{dg76@rice.edu}
\orcid{0009-0008-5786-755X}
\author{Laleh Aghababaie Beni}
\affiliation{Google Quantum AI, Venice, CA, 90291}
\email{lalehbeni@google.com}
\author{Noah Shutty}
\affiliation{Google Quantum AI, Venice, CA, 90291}
\email{shutty@google.com}

\begin{abstract}
Quantum Error Correction (\QEC{}) is essential for building robust, fault-tolerant quantum computers; however, the decoding process often presents a significant computational bottleneck. \Tesseract{} is a novel Most-Likely-Error (\textit{MLE}) decoder for \QEC{} that employs the \astar{} search algorithm to explore an exponentially large graph of error hypotheses, achieving high decoding speed and accuracy. This paper presents a systematic approach to optimizing the \Tesseract{} decoder through low-level performance enhancements. Based on extensive profiling, we implemented four targeted optimization strategies, including the replacement of inefficient data structures, reorganization of memory layouts to improve cache hit rates, and the use of hardware-accelerated bit-wise operations. We achieved significant decoding speedups across a wide range of code families and configurations, including Color Codes, Bivariate-Bicycle Codes, Surface Codes, and Transversal CNOT Protocols. Our results demonstrate consistent speedups of approximately 2$\times$ for most code families, often exceeding 2.5$\times$. Notably, we achieved a peak performance gain of over 5$\times$ for the most computationally demanding configurations of Bivariate-Bicycle Codes. These improvements make the \Tesseract{} decoder more efficient and scalable, serving as a practical case study that highlights the importance of high-performance software engineering in \QEC{} and providing a strong foundation for future research.
\end{abstract}

\section{Introduction}\label{sec:introduction}
Quantum computing represents a significant paradigm shift in high-performance computing, as quantum computers can solve problems that even the most powerful classical supercomputers cannot. Their immense power stems from \textbf{\textit{qubits}}, which, unlike classical bits restricted to states of 0 or 1, can exist in a superposition of both. This enables them to store and process significantly more information, but introduces a critical vulnerability: qubits are fragile and susceptible to noise and unwanted environmental interactions. These factors lead to a rapid accumulation of errors, such as decoherence and gate errors. To overcome this, Quantum Error Correction (\QEC{}) algorithms are executed on classical computers to constantly detect and fix errors on quantum machines. This prevents error accumulation over time and enables the construction of robust and fault-tolerant quantum machines.

The \QEC{} process operates by collecting \textbf{\textit{syndromes}}, which are measurements that indicate the presence and location of errors that occurred in the quantum system. The primary goal of a \QEC{} decoder is to identify the most probable error configuration consistent with the input measurement syndrome. This task can be computationally intensive, especially when decoding large and complex quantum codes, as decoders must explore a vast space of error configurations to identify the most probable one. Therefore, enhancing both the speed and accuracy of \QEC{} decoders is essential for building robust and fault-tolerant quantum computers.

This paper focuses on accelerating the decoding speed of several quantum code families: Surface Codes~\cite{surface1,surface2,surface3} (2D topological codes known for their high error thresholds and local qubit connectivity), Color Codes~\cite{color1,color2,color3,color4,color5} (a generalization of topological codes defined on tri-colorable lattices, offering higher encoding rates), Bivariate-Bicycle Codes~\cite{bicycle1,bicycle2} (a code family that leverages algebraic constructions to achieve high encoding rates), and Transversal CNOT Protocols for Surface Codes~\cite{trans1,trans2} (fault-tolerant methods for implementing logical CNOT gates on Surface Codes). Our work specifically improves the \Tesseract{}~\cite{tesseract} decoder, a search-based decoder that employs the \astar{}~\cite{astar} algorithm to efficiently explore an exponentially large graph of error configurations and identify the most probable configuration consistent with the input measurement syndrome. The original authors of \Tesseract{} have shown that this decoder can decode various quantum code families; in this paper, we introduce novel optimizations to significantly boost its decoding speed while maintaining its high accuracy.

\subsection{Major Contributions}
We present a systematic optimization of the \Tesseract{} decoder by identifying and addressing low-level computational bottlenecks through extensive profiling. Our work achieves significant performance gains across diverse code families without compromising decoding accuracy. The primary contributions of our work are as follows:

\begin{itemize}
    \item \textbf{Significant and Scalable Decoding Speedups:} We demonstrate consistent speedups of approximately 2$\times$ across most tested code families, frequently exceeding 2.5$\times$. Notably, we achieved a peak performance gain of over 5$\times$ for the most computationally demanding configurations of Bivariate-Bicycle codes, reducing the decoding time for 1,000 simulations from over 36,000 seconds to approximately 7,000 seconds.
    
    \item \textbf{Architectural Optimizations for Data Locality:} We identify and mitigate critical hardware-level bottlenecks, specifically addressing high cache miss rates by optimizing memory layouts for core heuristic functions. Furthermore, we eliminate the overhead of bit-packed data structures by utilizing hardware-accelerated bit-wise operations, significantly improving instruction-level parallelism.

    \item \textbf{Comprehensive Dataset and Step-by-Step Evaluation:} We provide a rigorous performance dataset across a vast parameter space. We employ a fine-grained, incremental evaluation strategy to quantify the individual impact of each optimization, providing an in-depth analysis of how specific code-level inefficiencies affect overall decoding throughput.
    
    \item \textbf{Robustness Across Diverse Code Families and Architectures:} Our evaluation spans a wide array of code families, including Surface Codes, Color Codes, Bivariate-Bicycle Codes, and Transversal CNOT Protocols. We validated these results across multiple noise models, various \astar{} search parameters, and three distinct hardware generations (Intel Xeon W-2135, Cascade Lake, and Sapphire Rapids).
\end{itemize}

\section{Related Work}\label{sec:related_work}
\QEC{} decoders operate by processing the input measurement syndrome and converting it into an internal representation suitable for decoding the most probable error configuration. These internal representations often utilize graph-based structures that model errors in the system and the relations between them. Identifying the most probable error configuration enables the correction of the error.

\textit{Minimum Weight Perfect Matching} (\textit{MWPM}) decoders~\cite{mwpm1,mwpm2} operate by converting the input measurement syndrome into a graph model where nodes are violated syndrome bits (called defects), and edges are potential physical error chains connecting defects. The goal of the decoder is to detect a minimum-weight perfect matching that connects nodes in the graph, which identifies the most probable error configuration consistent with the syndrome. \textit{MWPM} decoders achieve high accuracy but have limited computational efficiency, often scaling polynomially with the code size.

\textit{Belief Propagation} (\textit{BP}) decoders~\cite{bp1,bp2} operate on the code's Tanner graph by passing probabilistic "beliefs" between variable nodes (qubits or errors) and check nodes (syndrome measurements). \textit{BP} decoders achieve high computational efficiency, often scaling linearly with the code size, but have lower accuracy as they may not always find the optimal solution. To address this, \textit{Belief Propagation with Ordered Statistics Decoding} (\textit{BP+OSD}) has emerged as a hybrid approach. This method performs an initial pass with \textit{BP} and then employs \textit{OSD} post-processing, which explores the most unreliable bit flips to find a more accurate, often optimal, solution. The \textit{BP+OSD} approach imposes a trade-off between accuracy and efficiency, as it has lower computational efficiency compared to pure \textit{BP}.

\textit{Union Find} (\textit{UF}) decoders~\cite{uf1,uf2} operate by converting the input measurement syndrome into clusters that grow around detected syndrome defects and merge when they connect. This process can efficiently identify the most probable error configuration. \textit{UF} decoders provide a strong balance of accuracy and efficiency. Their computational efficiency often scales near-linearly with the code size, with logical error rates comparable to, or even surpassing, \textit{MWPM} decoders.

\textit{Correlated Matching} (\textit{CM}) decoders~\cite{cm1,cm2} explicitly account for error dependencies that arise from multi-qubit gates or device-specific interactions in a real quantum system. They operate by converting the input measurement syndrome into a more sophisticated model than the simple graphs used in prior work. They achieve this by introducing calibrated edge weights based on noise characterization or by employing hypergraph representations. \textit{CM} decoders can achieve high accuracy, pushing closer to theoretical error thresholds.

\textit{Integer Programming} (\textit{IP}) decoders~\cite{ip1,ip2,ip3} operate by converting the input measurement syndrome into an Integer Linear Program (ILP) and decoding the error configuration by solving the ILP. These decoders provide highly accurate and often optimal solutions, serving as a powerful benchmark for other algorithms. In this paper, we leveraged an \textit{IP} decoder to rigorously verify that the performance optimizations we implemented inside \Tesseract{} did not degrade its decoding accuracy when processing identical input measurement syndromes.

\section{Tesseract Decoder}\label{sec:tesseract}
\Tesseract{}~\cite{tesseract} is a novel Most-Likely-Error (\textit{MLE}) decoder for Low-Density-Parity-Check (\textit{LDPC}) quantum codes. \Tesseract{} finds the most likely error configuration by identifying the lowest-cost configuration consistent with the input measurement syndrome. Unlike many decoders that start with a polynomial-time algorithm and then employ heuristics to improve accuracy, \Tesseract{} starts with an exponential-time algorithm that identifies the most likely configuration and then employs heuristics to improve speed.

\Tesseract{} formulates the \textit{MLE} problem as a shortest path problem on an exponentially large graph of error configurations. In this graph, nodes represent configurations, and edges represent transitions formed by adding a single new error to a current configuration. \Tesseract{} employs the \astar{} algorithm to find the lowest-cost error configuration, which represents the optimal configuration for the given syndrome. \Tesseract{} guides the \astar{} search towards the optimal solution by calculating the cost of a search state using the \getdetcost{} function, which we analyze in detail in Section~\ref{sec:getdetcost1} and~\ref{sec:getdetcost2}, and by prioritizing states with lower costs.

The original authors of \Tesseract{} implemented several heuristic techniques to accelerate the decoder's speed by sacrificing some accuracy:

\begin{itemize}
    \item \textbf{\textit{Admissible \astar{} Heuristic}}: \Tesseract{} employs an admissible \astar{} heuristic function to calculate a strict lower bound on the cost of each search state. This property guides the \astar{} search towards the optimal solution, reaching the "EXIT" node—the state representing the error configuration consistent with the input measurement syndrome. The admissibility of the heuristic guarantees finding the optimal configuration.
    \item \textbf{\textit{Graph Pruning:}} \Tesseract{} prunes the search space by employing a predicate that restricts which errors can be added to the current error configuration. This effectively makes the search graph a tree and eliminates the exploration of redundant paths.
    \item \textbf{\textit{Beam Search:}} \Tesseract{} integrates beam search into the \astar{} search by employing a "beam cutoff" that prunes nodes with too many residual detection events, specifically those with a count of detection events significantly exceeding a minimum observed value. Beam search can significantly prune the search space and imposes a trade-off between accuracy and speed: smaller beam sizes yield higher efficiency but degrade accuracy more, as they risk finding a suboptimal solution if the optimal path is pruned.
    \item \textbf{\textit{No-revisit Detection Events and Residual Detection Penalty:}} \Tesseract{} employs a heuristic that avoids re-visiting search states with previously explored patterns of fired detectors. This heuristic can significantly prune the search space and prevent the exploration of redundant paths. In addition, \Tesseract{} can introduce a penalty cost for residual detection events to discourage high-cost paths.
    \item \textbf{\textit{At Most Two Errors Per Detector:}} \Tesseract{} can employ an optional heuristic that assumes at most two errors can affect any given detector. When enabled, this heuristic prevents adding errors to the current configuration that would result in more than two errors affecting any fired detector. This heuristic, when applied to a specific code family and/or noise model, can significantly prune the search space and accelerate decoding.
    \item \textbf{\textit{Priority Queue Size Management:}} \Tesseract{} utilizes a priority queue to manage search states and their associated costs, prioritizing those with lower costs to guide the search towards the optimal solution. To control memory usage and computational overhead, \Tesseract{} employs a heuristic that terminates the search if the priority queue size exceeds a predefined maximum number of states.
\end{itemize}

The authors of~\cite{tesseract} have benchmarked \Tesseract{} and demonstrated its performance advantages compared to other decoders. These benchmarks showed that \Tesseract{} is substantially faster than Integer Programming decoders while maintaining comparable accuracy at moderate physical error rates. For Surface, Color, and Bivariate-Bicycle Codes, \Tesseract{} and IP decoders achieved logical error rates one to two orders of magnitude lower than \textit{BP+OSD} decoders. Furthermore, \Tesseract{} is capable of decoding Transversal CNOT Protocols for Surface Codes on neutral atom quantum computers, protocols where other decoders often struggle with correlated errors. \Tesseract{} exhibits high decoding accuracy that enables a precise comparison of different quantum code families. For instance, the authors of~\cite{tesseract} revealed that Bivariate-Bicycle Codes can be $14\times$ to $19\times$ more efficient than Surface Codes when decoded with \Tesseract{}, whereas using \textit{CM} and \textit{BP+OSD} decoders implied only a $10\times$ improvement. The authors also compared \Tesseract{} with concurrently developed \textit{Decision-Tree-Decoders} (\textit{DTDs})~\cite{dtd} and showed that \Tesseract{} has superior performance.

In this paper, we focus on accuracy-preserving optimizations. Since \Tesseract{} was implemented in high-performance \Cpp{}, we delve deep into its open-source~\cite{tesseract-repo} implementation to investigate opportunities to improve decoding speed. Given that efficient software is crucial for building robust quantum computers, we identified subtle performance bottlenecks inside the decoder and implemented optimizations to address them.

\section{Analysis and Optimization}\label{sec:optimizations}
We utilized two profiling tools to measure \Tesseract{}'s performance, identify critical bottlenecks, and implement optimizations. Initially, we utilized \textit{HPCToolkit}~\cite{hpctoolkit-parallel, hpctoolkit-gpu}, a profiling tool developed at Rice University, to gather detailed, source-code-level performance insights. Its graphical user interface, which offers top-down, bottom-up, and flat views, was instrumental during the early phase of granular bottleneck identification. Subsequently, owing to its lower overhead and ease of use, we switched to \textit{perf}~\cite{perf} to effectively evaluate the performance gains achieved after implementing each optimization. Additionally, when evaluating each modification, we compared \Tesseract{}'s decoding output with the output produced by the Integer Programming decoder used as a benchmark in the original paper~\cite{tesseract}, ensuring that our optimizations did not compromise the decoder's accuracy.

\subsection{Addressing Bit-Packed Data Access Overheads}\label{sec:vector_bool}
Our initial profiling revealed significant performance overheads stemming from the extensive use of \vbool{} data structures. \Tesseract{} initially used these structures to store and manage patterns of fired detectors (representing the measurement syndrome) and predicates that block specific errors from being added to the current error configuration—a key graph pruning technique detailed in Section~\ref{sec:tesseract}. The issue with the \vbool{} data structure is that it is a specialized \Cpp{} template that packs elements as individual bits. While this minimizes the memory footprint, it introduces a significant overhead when elements are frequently accessed or modified. This occurs because individual elements are not directly addressable bytes; instead, accessing and modifying them requires the use of proxy objects that perform costly bit-wise operations, including shifting and masking.

The computational intensity of these bit-wise operations was further exacerbated by \Tesseract{}'s core heuristics. \Tesseract{}'s graph pruning heuristic involves frequently blocking and unblocking errors within \vbool{}. Furthermore, the "No-Revisit Detection Events" heuristic, which avoids re-visiting search states with previously explored patterns of fired detectors, requires hashing these patterns. This process necessitates iterating through each element of the \vbool{} containing the pattern. These frequent access and modify operations across numerous decoding iterations created significant overheads in code regions operating with \vbool{}.

\begin{figure*}
  \centering
  \includegraphics[width=.7\textwidth]{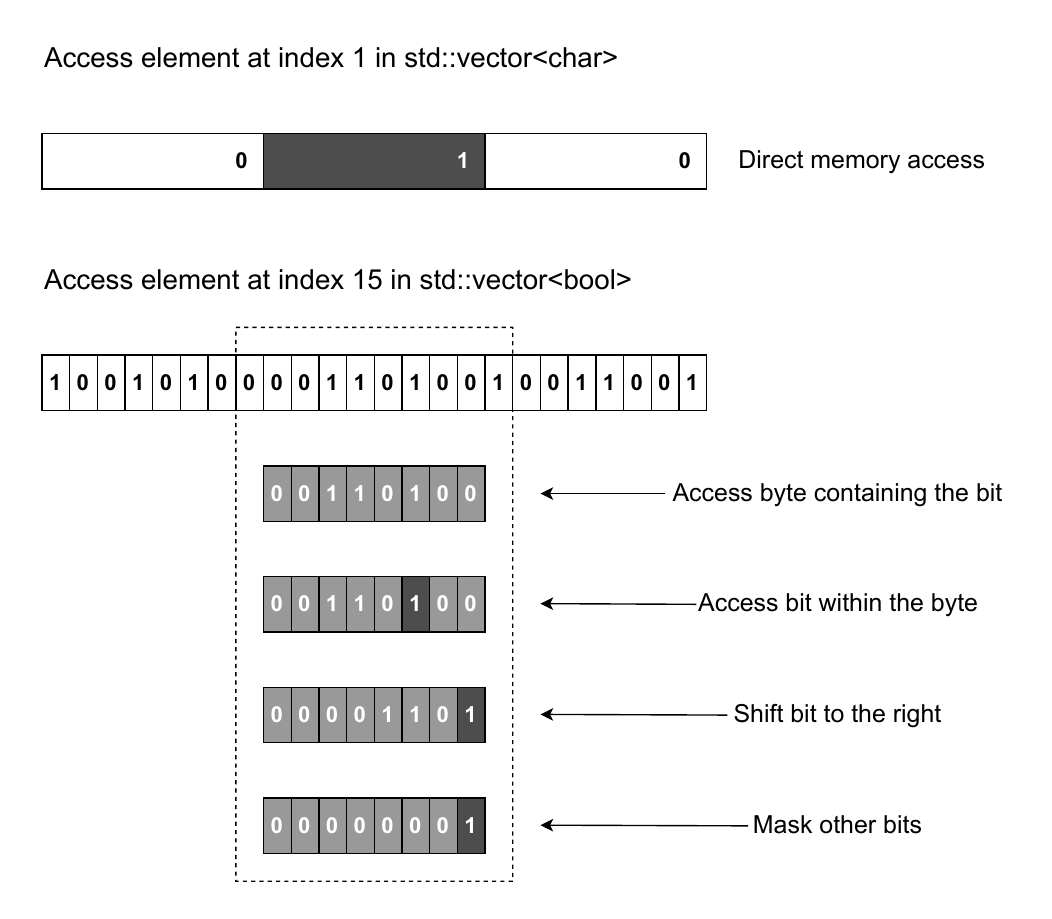}
  \caption{Comparison of memory access patterns between \vbool{} and \vchar{}. While \vbool{} minimizes memory footprint through bit-packing, frequent access requires proxy objects to perform bit-wise shifting and masking, inducing significant computational overhead compared to direct byte-addressable access in \vchar{}.}
  \label{fig:vector_bool}
\end{figure*}

Our initial approach to address these inefficiencies was to systematically replace instances of \vbool{} with \vchar{}. While this increased the memory footprint by storing individual elements as full bytes, our analysis revealed that for \Tesseract{}'s specific workload, the performance bottleneck was dominated by the high overhead of bit-wise operations inside \vbool{} rather than memory copying or overall memory consumption. Therefore, the performance gains achieved by enabling direct byte-level access with \vchar{} significantly outweighed any increased memory consumption. This optimization led to significant decoding speedups across various code families and configurations, which we analyze in detail in Section~\ref{sec:results_short_beam}. Fig.~\ref{fig:vector_bool} illustrates the memory access patterns in \vbool{} and \vchar{}.

\subsection{Cache-Optimized Data Reorganization}\label{sec:getdetcost1}
As detailed in Section~\ref{sec:tesseract}, \Tesseract{} employs an admissible \astar{} heuristic function to guide the search towards the optimal solution. This is achieved by calculating a strict lower bound on the cost of each search state and prioritizing states with lower costs. \Tesseract{} computes the cost of a search state using the \getdetcost{} function, which aggregates the minimum cost of all errors affecting a specific detector, provided those errors are not blocked in the current decoding iteration due to graph pruning rules. Listing~\ref{lst:get_detcost_before} shows the initial implementation of the \getdetcost{} function. Each time \Tesseract{} considers exploring a search state—by adding a new error to the current error configuration—it first calculates the cost of that state. This involves computing the cost of each detector (via \getdetcost{}) potentially affected by the new error, as well as detectors affected by errors related to the current error configuration. While this process of calculating an accurate state cost is essential for guiding the search, it represents a major performance bottleneck.

\begin{figure}
\centering
\begin{lstlisting}[
    caption={Initial implementation of the \getdetcost{} function. This function calculates the minimum error cost for a given detector \texttt{d} by iterating through the list of potential errors \texttt{d2e[d]} that are not currently blocked.},
    label=lst:get_detcost_before
]
double TesseractDecoder::get_detcost(size_t d,
                                    const std::vector<bool>& blocked_errs,
                                    const std::vector<size_t>& det_counts) {
  double min_cost = INF;
  for (size_t ei : d2e[d]) {
    if (!blocked_errs[ei]) {
      double ecost = errors[ei].likelihood_cost / det_counts[ei];
      min_cost = std::min(min_cost, ecost);
    }
  }
  return min_cost + config.det_penalty;
}
\end{lstlisting}
\end{figure}

The primary inefficiency within the \getdetcost{} function stemmed from its frequent access to elements at arbitrary, non-contiguous indices within two large pre-computed vectors: one indicating currently blocked errors and another storing the total number of fired detectors for each error. The function accesses errors affecting the selected detector inside these vectors, computes their cost (if not blocked), and determines the detector's minimum error cost. These high-frequency accesses to scattered memory locations within two large vectors led to numerous cache misses, severely degrading performance.

\begin{figure*}
  \centering
  \includegraphics[width=\textwidth]{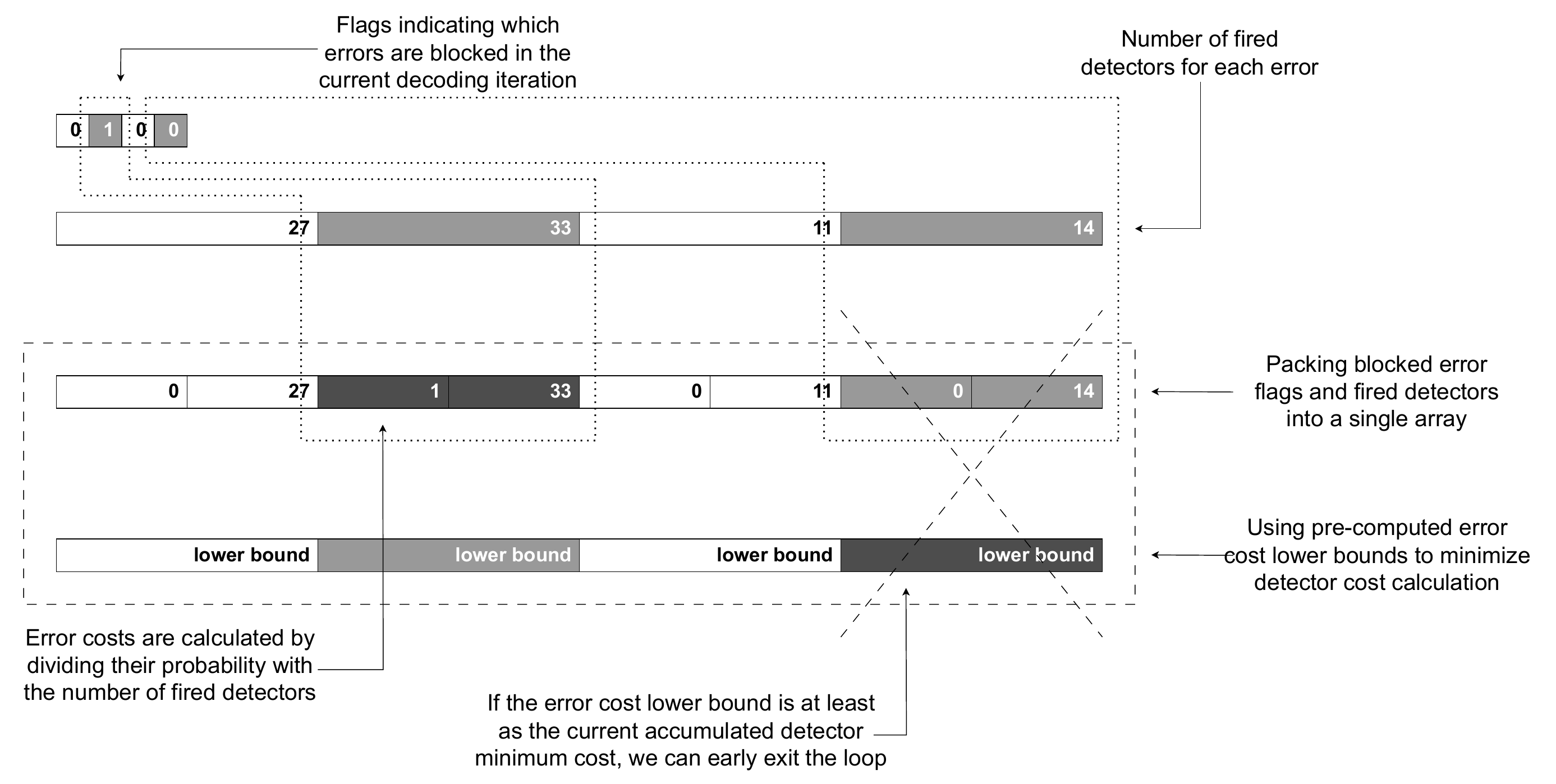}
  \caption{Data reorganization and early-exit logic within the \getdetcost{} function. Consolidating frequently accessed elements from separate vectors into a single contiguous structure (Array of Structs) optimizes CPU cache performance. The integrated early-exit strategy utilizes pre-calculated error cost lower bounds to bypass unnecessary computations.}
  \label{fig:get_detcost}
\end{figure*}

A key observation in the initial implementation was that while accesses to these two large vectors were at arbitrary indices, they exhibited a consistent co-access pattern: elements at the same index in both vectors were always accessed together. To address the cache inefficiency and leverage this concurrent access pattern, we redesigned these two vectors into a single vector containing custom \texttt{DetectorCostTuple} data structures. This technique is similar to converting a Structure of Arrays (SoA) to an Array of Structs (AoS)~\cite{SoAvsAoS}. The \texttt{DetectorCostTuple} structure contains two \texttt{uint32\_t} (4-byte unsigned integer) fields: one for the blocked error flag and the other for the fired detectors count. This reorganization dramatically improved data locality, ensuring that co-accessed data resided contiguously in memory. Furthermore, by designing the structure with a total size that aligns well with CPU cache lines (typically 64 bytes on most modern architectures), we maximized the benefits of hardware pre-fetching. Fig.~\ref{fig:get_detcost} illustrates the data reorganization performed inside the \getdetcost{} function. This optimization had the highest impact on code families and configurations that spent significant time in the \getdetcost{} function, such as Color Codes and Bivariate-Bicycle Codes.

\subsection{Early-Exit Strategy for Minimizing Cost Calculation}\label{sec:getdetcost2}
Listing~\ref{lst:get_detcost_before} shows that the cost of a detector is computed as the minimum cost among errors that can affect it and are not blocked. Costs of individual errors are calculated by dividing their \texttt{likelihood\_cost} (constant throughout the decoding process) by the number of detectors fired by that error in the current iteration. The \getdetcost{} function iterates through every error affecting the selected detector, computes its cost, and identifies the minimum.

We realized that we could significantly prune iterations within the \getdetcost{} loop by computing lower bounds for error costs. While the number of detectors fired by an error changes across iterations, each error has a maximum possible number of detectors it can affect, and the number of fired detectors will never exceed that maximum. This maximum provides a baseline for a lower bound: since a smaller denominator yields a larger cost, the largest number of fired detectors gives the smallest possible (lower bound) cost for an error.

Therefore, before the decoding starts, we pre-compute lower bounds for error costs by dividing their \texttt{likelihood\_cost} by the maximum possible number of fired detectors. We store these pre-computed lower bounds in an array and, for each detector, sort its relevant errors based on these lower bounds. This enables an effective early-exit strategy within the \getdetcost{} loop: if the currently found \texttt{min\_cost} is less than or equal to the lower bound of the next error to be evaluated, further iterations are unnecessary, as no subsequent error can yield a smaller cost due to the sorted order. Listing~\ref{lst:get_detcost} shows the implementation of the \getdetcost{} function after optimizations. Fig.~\ref{fig:get_detcost} also illustrates the logic of using pre-computed error cost lower bounds to avoid unnecessary computations. This optimization had the highest impact on code families that spent significant time in the \getdetcost{} function, similar to the second optimization.

\begin{figure}
\centering
\begin{lstlisting}[caption={Optimized \getdetcost{} implementation. This version leverages a unified \texttt{DetectorCostTuple} structure (Array of Structs) to improve cache locality and utilizes pre-sorted error cost lower bounds to facilitate an early-exit strategy.}, label=lst:get_detcost]
double TesseractDecoder::get_detcost(size_t d, const std::vector<DetectorCostTuple>& detector_cost_tuples) {
  double min_cost = INF;
  ErrorCost ec;
  DetectorCostTuple dct;

  for (size_t ei : d2e[d]) {
    ec = error_costs[ei];
    if (ec.min_cost >= min_cost) break;

    dct = detector_cost_tuples[ei];
    if (!dct.error_blocked) {
      double error_cost = ec.likelihood_cost / dct.detectors_count;
      min_cost = std::min(min_cost, error_cost);
    }
  }

  return min_cost + config.det_penalty;
}
\end{lstlisting}
\end{figure}

\subsection{Efficient Hashing of Syndrome Patterns}\label{sec:hashing}
After implementing and evaluating the first three optimizations, we identified another opportunity for improvement. Following the second optimization, where we reorganized data consumed by \getdetcost{} into a custom \texttt{DetectorCostTuple} structure, the only remaining \vchar{} instance was the one storing patterns of fired detectors (representing the measurement syndrome). We observed that the function responsible for hashing these patterns consumed a significant portion of the decoding time. This function is a critical component of the "No-revisit Detection Events" heuristic, detailed in Section~\ref{sec:tesseract}, which stores previously explored patterns in a \texttt{std::set} to prevent redundant computations. The hashing function was initially implemented as a simple loop that iterates through each element of the vector and accumulates the hash (Listing~\ref{lst:hashing}). While the switch from \vbool{} to \vchar{} provided speedups, this function still consumed a significant amount of decoding time, especially for code families with bushy search graphs where the \astar{} search generates more redundant paths.

\begin{figure}
\centering
\begin{lstlisting}[caption={Unoptimized hashing implementation for syndrome patterns, utilizing a standard accumulation loop over \texttt{std::vector<char>}.}, label=lst:hashing]
struct VectorCharHash {
  size_t operator()(const std::vector<char>& v) const {
    size_t seed = v.size();

    for (char el : v) {
      seed = seed * 31 + static_cast<size_t>(el);
    }
    return seed;
  }
};
\end{lstlisting}
\end{figure}

We sought a data structure that would provide a highly optimized, built-in hashing function. The \Cpp{} standard library provides \texttt{std::bitset}, which combines memory-efficient bit-packing with hardware-accelerated bit-wise operations. This is achieved by internally storing bits within contiguous memory blocks that can execute vectorized bit-wise operations. However, \texttt{std::bitset} requires a static size determined at compile-time, making it unsuitable for \Tesseract{}'s dynamically sized syndrome patterns. This led us to explore alternatives that could provide a dynamically sized data structure with similar hardware-accelerated operations.

We turned to the \textit{Boost} library~\cite{boost}, which provides \texttt{boost::dynamic\_bitset}. Similar to \texttt{std::bitset}, this data structure internally stores bits within contiguous memory blocks that enable the execution of hardware-accelerated bit-wise operations, thereby avoiding the overhead of proxy objects present in \vbool{}. Additionally, \texttt{boost::dynamic\_bitset} supports dynamically sized arrays. We adopted \texttt{boost::dynamic\_bitset} to store \Tesseract{}'s syndrome patterns and used the built-in \texttt{boost::hash\_value} function to accelerate the hashing. This approach not only mitigated the increased memory footprint from our initial optimization with \vchar{} but also provided a highly optimized method for hashing boolean sequences, further enhancing decoding speed. This optimization had the highest impact on Surface Codes and Transversal CNOT Protocols, especially for \astar{} with longer beam sizes, where search graphs contain many redundant paths that must be checked.

\section{Results}\label{sec:results}
This section details the performance gains achieved through a systematic optimization of the \Tesseract{} decoder. Our evaluation spans a diverse range of quantum code families—including Color Codes, Bivariate-Bicycle Codes, Surface Codes, and Transversal CNOT Protocols—under varied noise models and search configurations. To demonstrate the portability and robustness of our solution across different architectural and search environments, we conducted benchmarks on three distinct hardware platforms using varying search parameters. For all tests, \Tesseract{} was compiled with consistent \texttt{-O3} flags to enable aggressive compiler optimizations; this ensures that the reported speedups reflect fundamental algorithmic and architectural improvements that standard compiler passes cannot achieve.

For each hardware architecture and search configuration, we executed a statistically significant volume of decoding simulations (shots), leveraging multi-threading to accelerate data collection. However, to provide a clear measure of algorithmic efficiency and eliminate confounding factors, all reported execution times represent the cumulative CPU time of individual, single-threaded decoding operations. Table~\ref{tab:optimizations} summarizes the four key optimization strategies implemented to accelerate the \Tesseract{} decoder.

\newcolumntype{L}{>{\raggedright\arraybackslash}X}
\begin{table}[ht]
    \small
    \caption{Optimization strategies for accelerating the \Tesseract{} decoder.}
    \label{tab:optimizations}
    \centering
    \renewcommand{\arraystretch}{1.4}
    \begin{tabularx}{\textwidth}{l | L}
        \toprule
        \textbf{Optimization} & \textbf{Technical Method \& Rationale} \\
        \midrule
        
        \textbf{Optimization 1} & Replace \vbool{} with \vchar{} to eliminate the overhead of bit-wise masking/shifting. \\
        
        \textbf{Optimization 2} & Transition from Structure of Arrays (SoA) to Array of Structures (AoS) to mitigate frequent cache misses inside \getdetcost{}. \\
        
        \textbf{Optimization 3} & Compute lower bounds for error costs to enable an early-exit strategy inside \getdetcost{}. \\
        
        \textbf{Optimization 4} & Vectorize hashing of syndrome patterns using \texttt{boost::dynamic\_bitset} to accelerate graph exploration. \\
        
        \bottomrule
    \end{tabularx}
\end{table}

\subsection{Short Beam Experiments}\label{sec:results_short_beam}
We conducted the initial set of experiments on a dedicated server equipped with an Intel(R) Xeon(R) W-2135 CPU @ 3.70GHz (6 cores, 12 threads), running Debian GNU/Linux. For this architecture, the \Tesseract{} \astar{} search was configured with a "beam cutoff" of 15 and a priority queue limit of 200,000, utilizing the "No-Revisit Detection Events" heuristic (detailed in Section~\ref{sec:tesseract}). Benchmarks were executed across the following configurations:

\begin{itemize}
    \item \textbf{Color Codes:} We evaluated the \textit{uniform} and \textit{SI1000}~\cite{si1000} noise models. For each, we tested error rates ($p$) of 0.001 and 0.002 across code distances ($d$) of 7, 9, and 11.
    \item \textbf{Bivariate-Bicycle Codes:} We utilized the \textit{SI1000} model along with two variants described in~\cite{tesseract}: the \textit{NLR5} and \textit{NLR10} noise models. These variants apply $5\times$ and $10\times$ higher noise strength to the two-qubit depolarizing channels following long-range CZ gates, respectively. We tested error rates of 0.001 and 0.002 at code distances of 6 and 10.
    \item \textbf{Surface Codes and Transversal CNOT Protocols:} Using the \textit{SI1000} noise model, we tested error rates of 0.001 and 0.002 for code distances of 7, 9, and 11.
\end{itemize}

This architecture served as the primary environment for profiling the \Tesseract{} decoder, identifying critical inefficiencies, and implementing targeted optimizations. By analyzing execution bottlenecks within these specific code configurations, we developed the optimizations described in the previous section and performed an initial evaluation of the resulting speedups.

For this short beam configuration, we performed a fine-grained analysis of decoding speedups by applying the optimizations listed in Table~\ref{tab:optimizations} incrementally. This step-by-step approach enabled us to quantify the cumulative impact of each modification on overall performance. For each code family and configuration, we executed 1,000 decoding simulations and measured the total execution time at each stage of the optimization process.

\begin{figure*}
\centering
\subfloat[\centering]{\includegraphics[width=.5\textwidth]{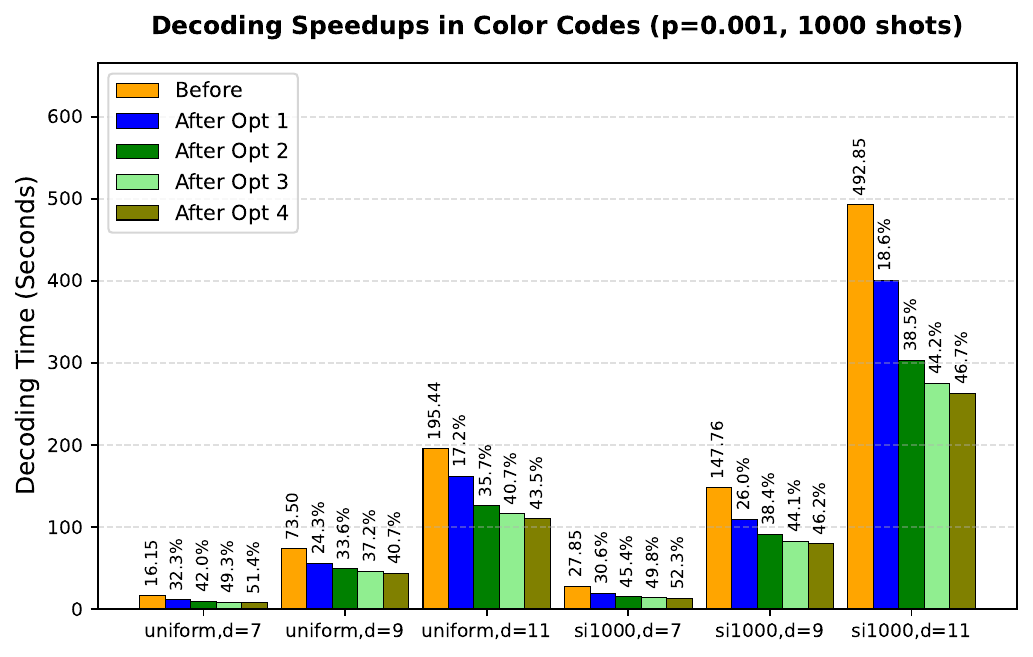}}
\subfloat[\centering]{\includegraphics[width=.5\textwidth]{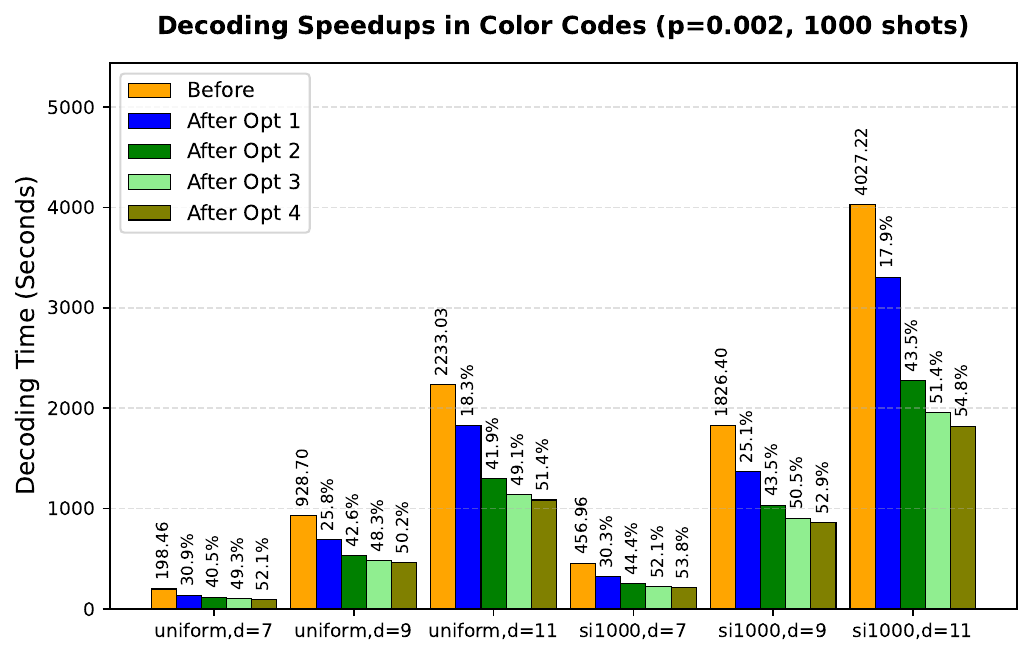}}\\
\subfloat[\centering]{\includegraphics[width=.5\textwidth]{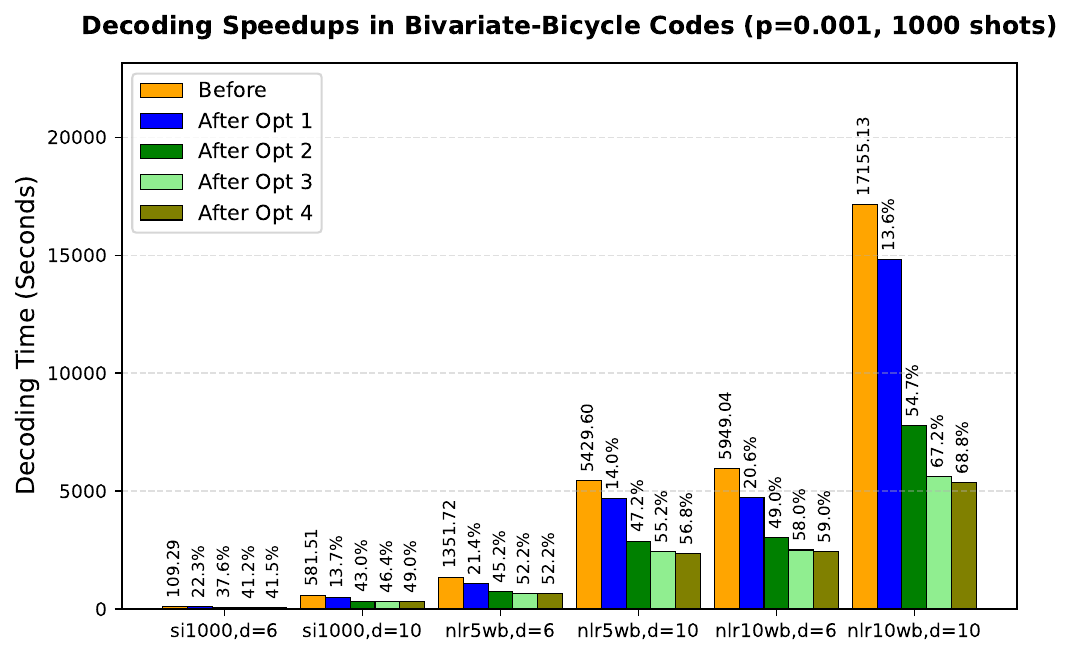}}
\subfloat[\centering]{\includegraphics[width=.5\textwidth]{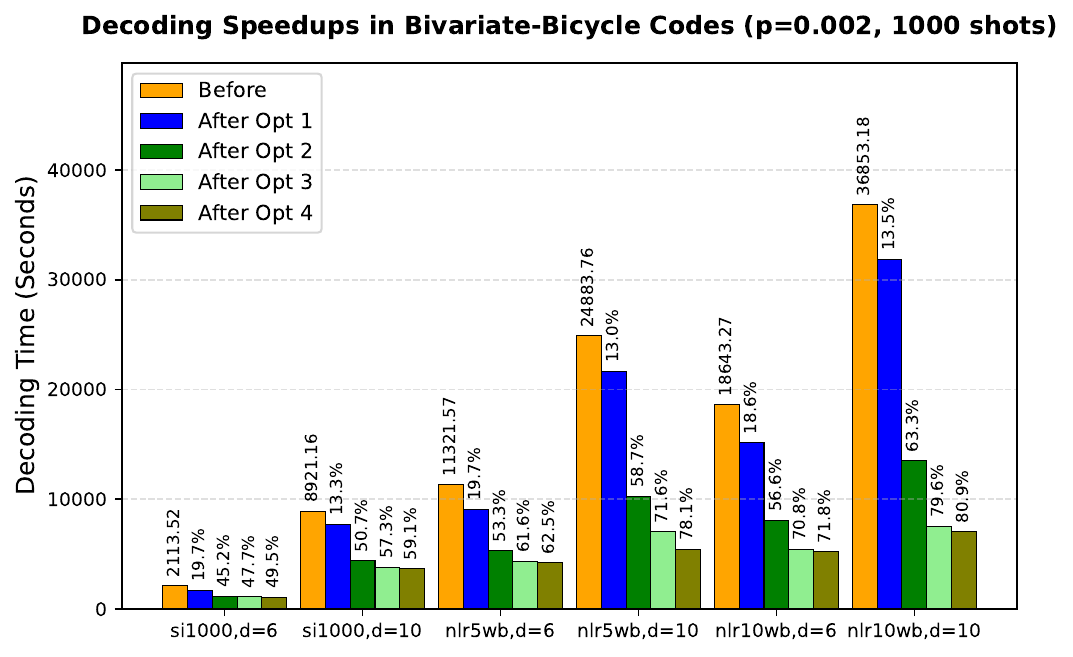}}\\
\subfloat[\centering]{\includegraphics[width=.5\textwidth]{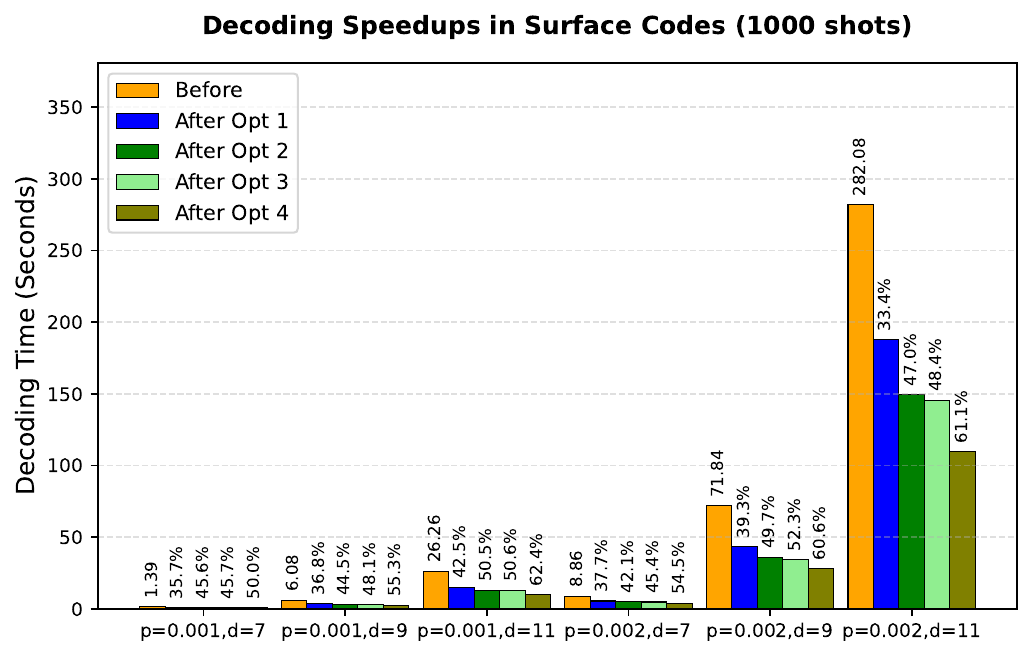}}
\subfloat[\centering]{\includegraphics[width=.5\textwidth]{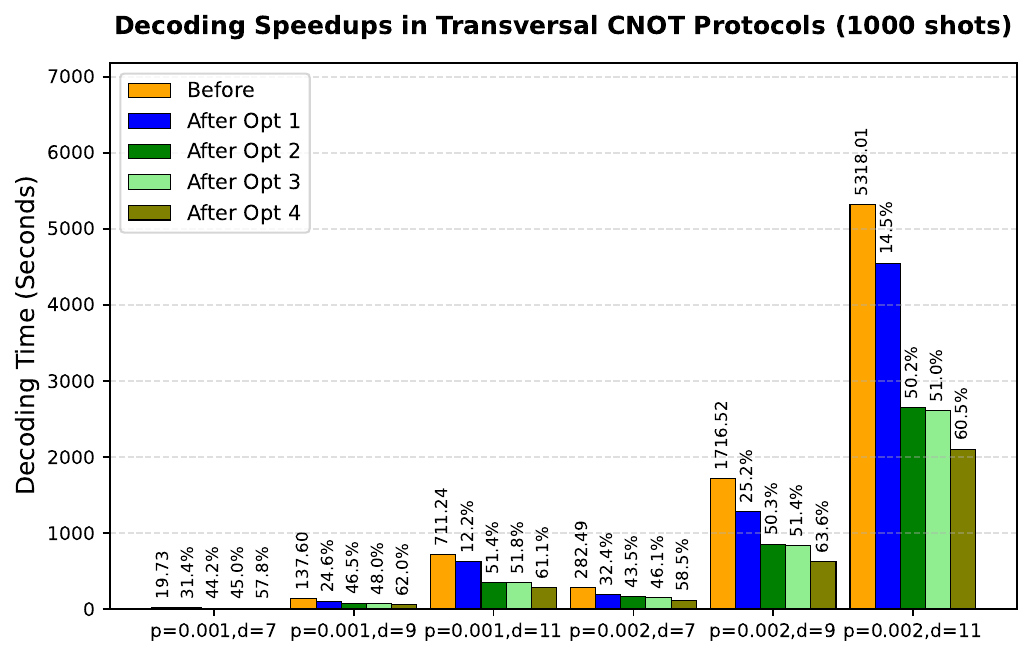}}
\caption{Decoding speedups across various code families and configurations for experiments on the first architecture using the short beam configuration: (\textbf{a}) Color Codes at $p=0.001$; (\textbf{b}) Color Codes at $p=0.002$; (\textbf{c}) Bivariate-Bicycle Codes at $p=0.001$; (\textbf{d}) Bivariate-Bicycle Codes at $p=0.002$; (\textbf{e}) Surface Codes; (\textbf{f}) Transversal CNOT Protocols for Surface Codes.}
\label{fig:google_results}
\end{figure*}

The results, presented in Figure~\ref{fig:google_results}, demonstrate that all four optimizations consistently and significantly reduced decoding times across all tested configurations. Notably, the effectiveness of these optimizations is most pronounced at higher error rates and larger code distances—the scenarios representing the most computationally demanding decoding tasks.

\textbf{\textit{Color Codes}}. Decoding latency for Color Codes increased significantly with both error rate and code distance. The implementation of \optone{} yielded an initial speedup of 17.2\%--32.3\%. Cumulative speedups after applying \opttwo{}, \optthree{}, and \optfour{} reached 33.6\%--45.4\%, 37.2\%--52.1\%, and 40.7\%--54.8\%, respectively. The most substantial gain occurred under the \textit{SI1000} noise model with $p=0.002$ and $d=11$, where the total speedup reached 2.21$\times$.

\textbf{\textit{Bivariate-Bicycle Codes}}. Bivariate-Bicycle Codes were the most computationally intensive, with decoding times exceeding 35,000 seconds for the most demanding configuration ($p=0.002$, $d=10$). \optone{} provided an initial speedup of 13.0\%--22.3\%, while the most dramatic performance gains followed the application of \opttwo{}, bringing cumulative speedups to 37.6\%--63.6\%. With the addition of \optthree{} and \optfour{}, cumulative speedups reached 41.2\%--79.6\% and 41.5\%--80.9\%, respectively. The maximum speedup was achieved for the \textit{NLR10} noise model at $p=0.002$ and $d=10$, reaching 5.24$\times$ and reducing the total execution time from 36,853.18 seconds to 7,048.21 seconds.

\textbf{\textit{Surface Codes}}. Surface Codes exhibited the lowest decoding latency among the tested families. \optone{} provided an initial speedup of 33.4\%--42.5\%. Cumulative speedups after \opttwo{}, \optthree{}, and \optfour{} reached 42.1\%--50.5\%, 45.4\%--52.3\%, and 50.0\%--62.4\%, respectively. The largest improvement was recorded for the configuration with $p=0.001$ and $d=11$, where the total speedup reached 2.66$\times$.

\textbf{\textit{Transversal CNOT Protocols}}. Decoding times for Transversal CNOT Protocols scaled sharply with higher error rates and larger code distances. \optone{} provided an initial speedup of 12.2\%--32.4\%. Cumulative speedups after \opttwo{}, \optthree{}, and \optfour{} reached 44.2\%--51.4\%, 45.0\%--51.8\%, and 57.8\%--63.6\%, respectively. The peak performance gain was achieved for the configuration with $p=0.002$ and $d=9$, resulting in a total speedup of 2.75$\times$.

These results underscore the substantial performance gains achievable through targeted code-level optimizations. Speedups exceeded $2\times$ for Color Codes, $2.5\times$ for Surface Codes and Transversal CNOT Protocols, and $5\times$ for the most computationally demanding Bivariate-Bicycle Code configurations. This demonstrates the immense potential for accelerating decoding simply by optimizing memory-access patterns and data organization.

The most impactful performance gains were derived from \opttwo{}, which transitioned the \getdetcost{} \astar{} heuristic function from a Structure of Arrays (SoA) to an Array of Structures (AoS) format. This modification proved critical, as the \getdetcost{} function initially accounted for over 50\% of the total decoding time and suffered from frequent cache misses. On this architecture, \opttwo{} contributed to a cumulative speedup of $2.75\times$ for the most demanding Bivariate-Bicycle configurations.

\optthree{} further addressed the \getdetcost{} bottleneck by implementing an early-exit strategy based on pre-computed error cost lower bounds, significantly pruning iterations within the function's primary loop. Consequently, \opttwo{} and \optthree{} were most effective for configurations where the heuristic function was the primary bottleneck—specifically in Bivariate-Bicycle Codes, where it consumed between 70\% and 90\% of the total execution time.

\optone{}, which replaced bit-vector representations with character vectors, provided a substantial initial speedup of up to 42\% in Surface Codes. Finally, \optfour{} vectorized the hashing of syndrome patterns, which also served to mitigate the increased memory footprint introduced by \optone{}. In our benchmarks on this architecture, \optfour{} had the most pronounced effect on Surface Codes and Transversal CNOT Protocols; their noise models, combined with the short beam configuration, produced "bushy" search graphs where the \astar{} search generates numerous redundant paths. Filtering these paths relies heavily on the hashing function, making vectorization particularly effective.

In the following section, we evaluate the performance gains across two additional hardware architectures using a larger beam search to assess the portability and scalability of our solution across different platforms and search configurations.

\subsection{Long Beam Experiments}\label{sec:results_long_beam}
As detailed in the previous section, the initial profiling and incremental evaluation of the \Tesseract{} decoder were performed on the first architecture using a short beam configuration. To assess the portability and scalability of our optimizations across different platforms and search settings, we conducted additional experiments on two high-performance architectures within Rice University's NOTS cluster~\cite{nots}. The first system is an Intel(R) Xeon(R) Gold 6230 CPU @ 2.10GHz (40 cores, 80 threads) based on the Cascade Lake microarchitecture, and the second is an Intel(R) Xeon(R) Platinum 8468 CPU @ 2.10GHz (96 cores, 192 threads) based on the Sapphire Rapids microarchitecture. Both systems run Red Hat Enterprise Linux 9.4.

For these experiments, we utilized a more extensive search configuration with a "beam cutoff" of 20. To balance the computational load of the larger beam—which explores a broader space of error hypotheses and converges more slowly—we set the priority queue limit to 100,000. This smaller memory threshold ensures that the most complex code configurations do not consume excessive time in the expanded search space. As with the previous experiments, we utilized the "No-Revisit Detection Events" heuristic and executed benchmarks across the same code configurations to ensure comparability:

\begin{itemize}
    \item \textbf{Color Codes:} Evaluated under \textit{uniform} and \textit{SI1000} noise models at error rates ($p$) of 0.001 and 0.002, with code distances ($d$) of 7, 9, and 11.
    \item \textbf{Bivariate-Bicycle Codes:} Evaluated under \textit{SI1000}, \textit{NLR5}, and \textit{NLR10} noise models at error rates of 0.001 and 0.002, with code distances of 6 and 10.
    \item \textbf{Surface Codes and Transversal CNOT Protocols:} Evaluated under the \textit{SI1000} noise model with error rates of 0.001 and 0.002 and code distances of 7, 9, and 11.
\end{itemize}

\begin{figure*}
\centering
\subfloat[\centering]{\includegraphics[width=.5\textwidth]{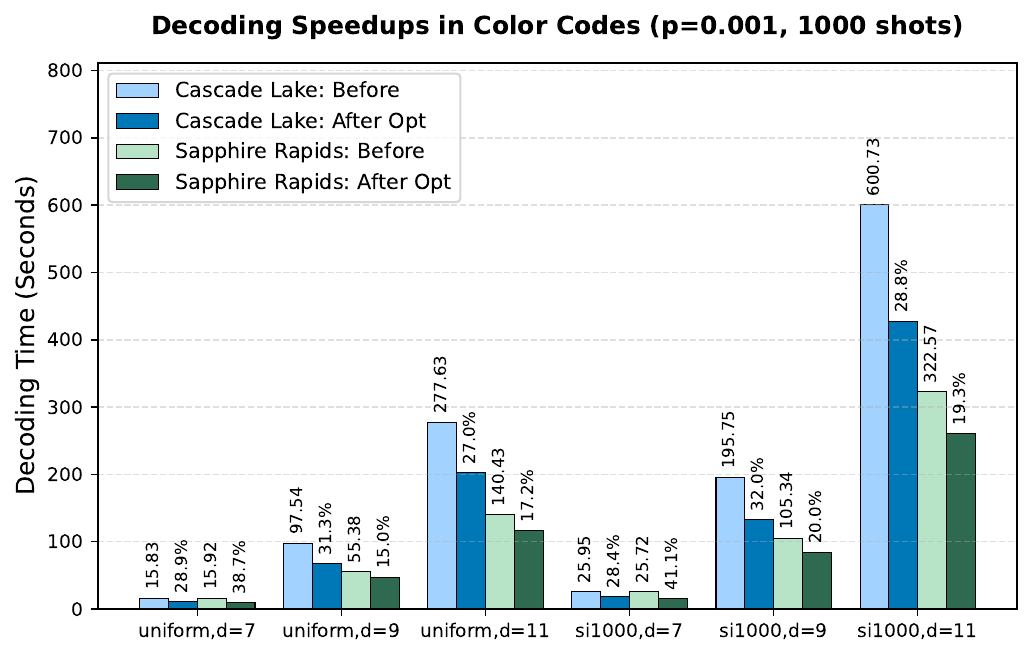}}
\subfloat[\centering]{\includegraphics[width=.5\textwidth]{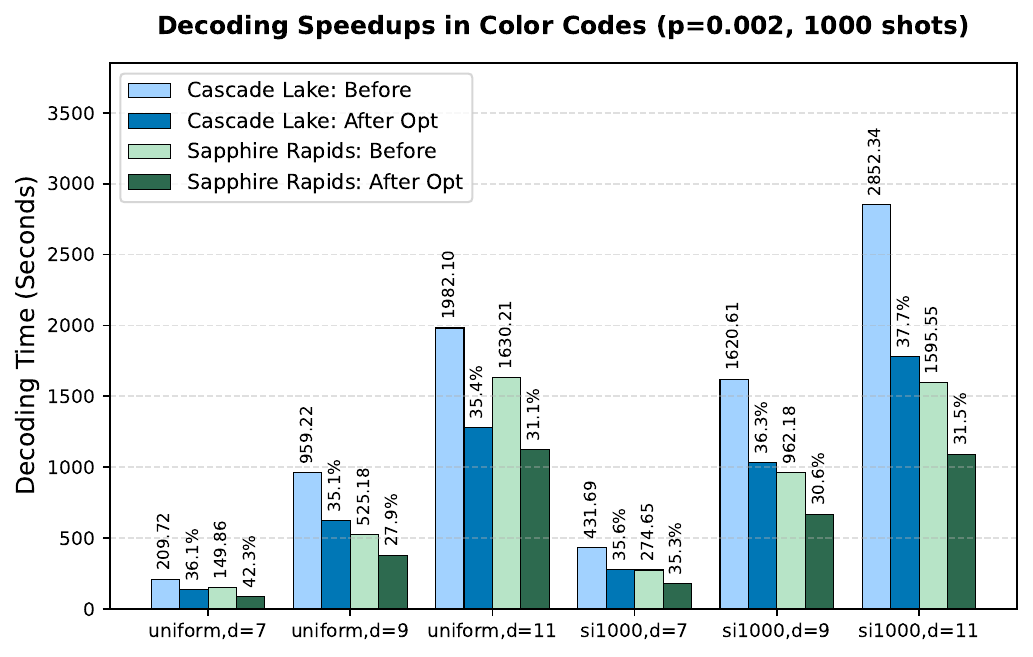}}\\
\subfloat[\centering]{\includegraphics[width=.5\textwidth]{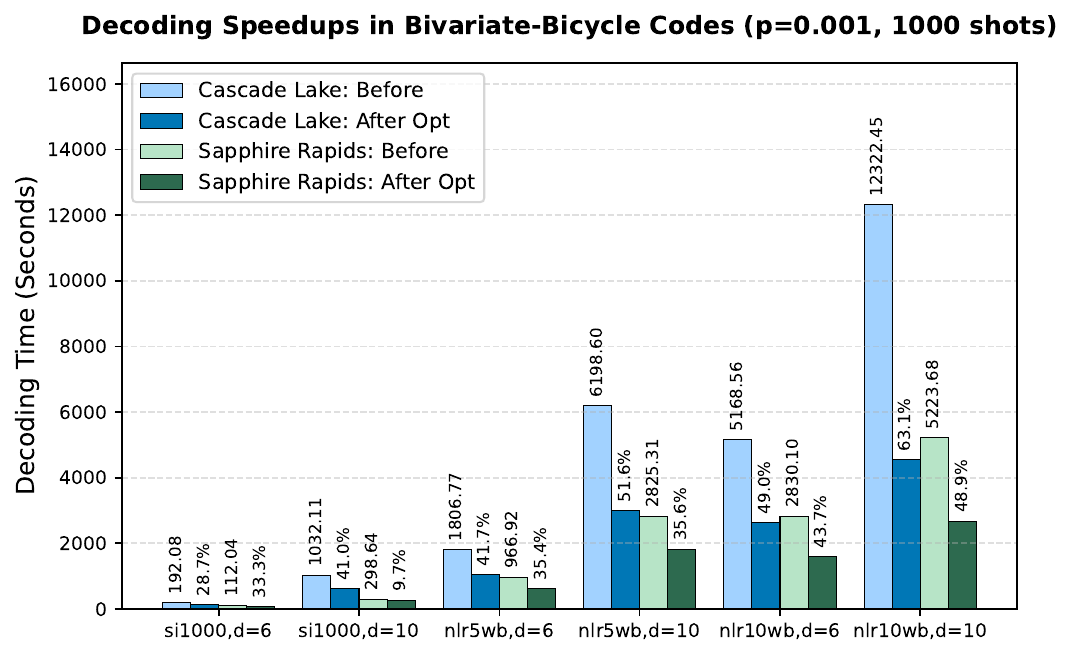}}
\subfloat[\centering]{\includegraphics[width=.5\textwidth]{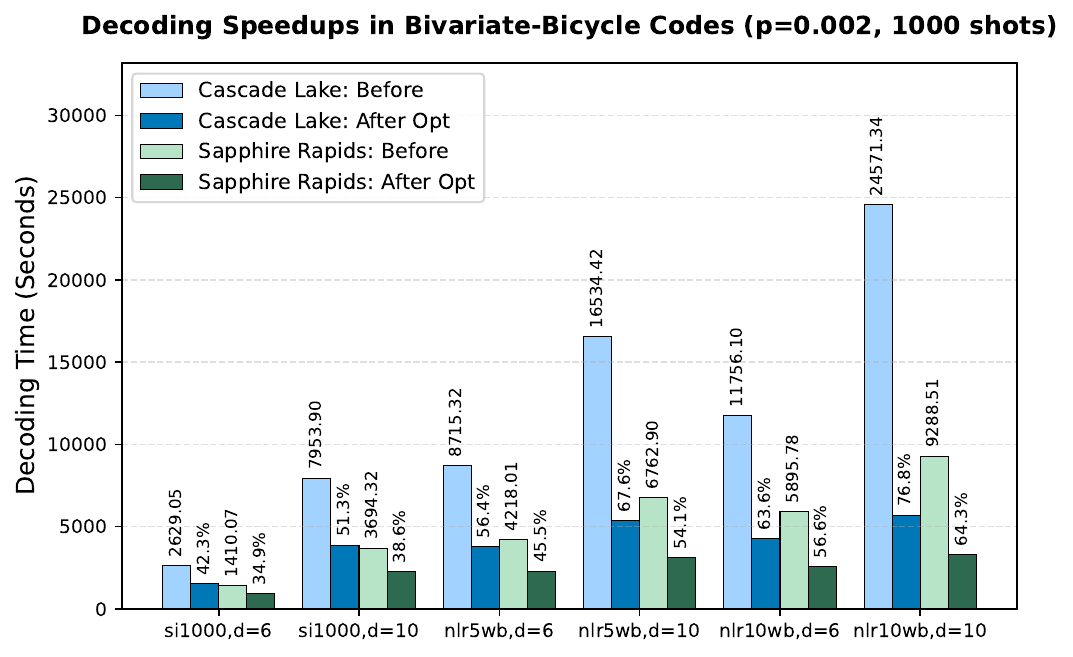}}\\
\subfloat[\centering]{\includegraphics[width=.5\textwidth]{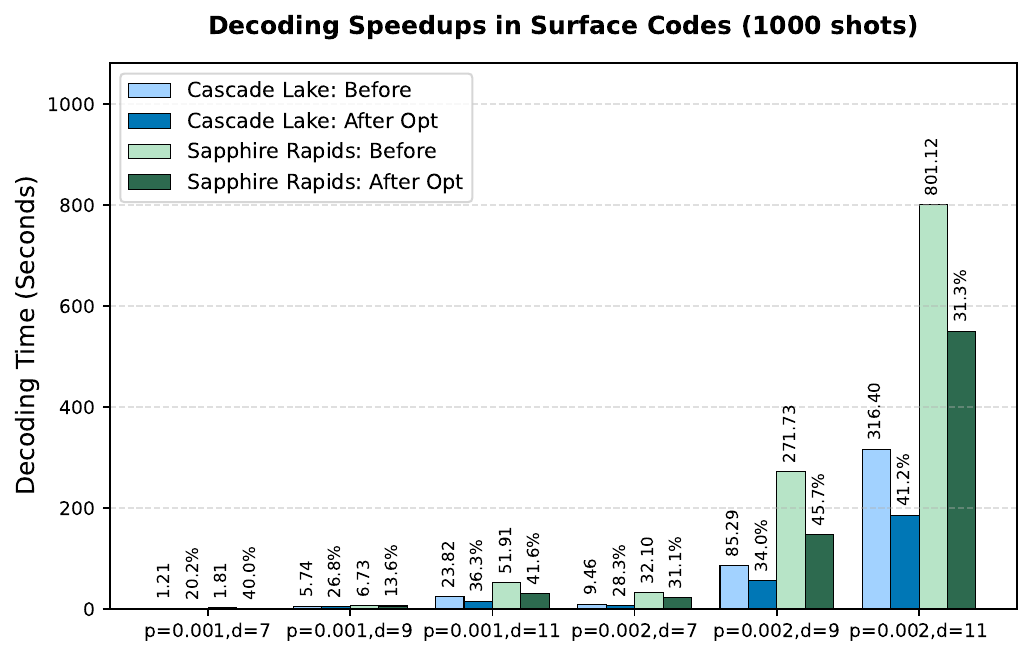}}
\subfloat[\centering]{\includegraphics[width=.5\textwidth]{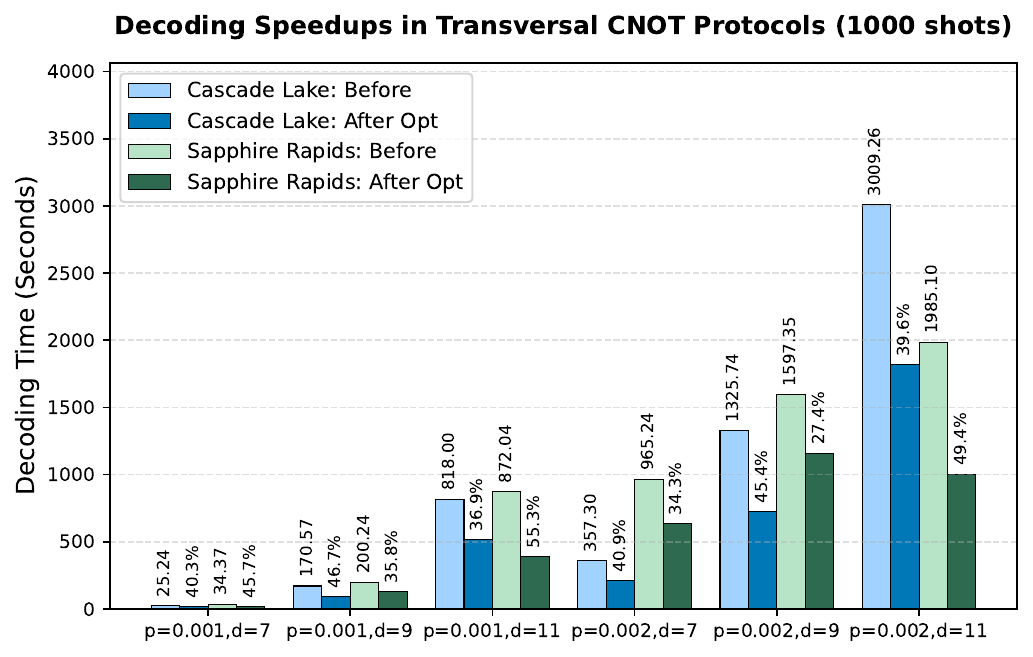}}
\caption{Decoding speedups achieved across various code families and configurations for the second (Cascade Lake) and third (Sapphire Rapids) architectures, utilizing the long beam configuration: (\textbf{a}) Color Codes at $p=0.001$; (\textbf{b}) Color Codes at $p=0.002$; (\textbf{c}) Bivariate-Bicycle Codes at $p=0.001$; (\textbf{d}) Bivariate-Bicycle Codes at $p=0.002$; (\textbf{e}) Surface Codes; (\textbf{f}) Transversal CNOT Protocols for Surface Codes.}
\label{fig:rice_results}
\end{figure*}

Consistent with our initial methodology, we executed 1,000 decoding simulations for each configuration. On these NOTS cluster nodes, we measured execution times for the baseline \Tesseract{} decoder and for the fully optimized version incorporating all strategies described in Section~\ref{sec:optimizations}. This enabled us to calculate the total speedup achieved on each architecture and compare these results with the peak improvements observed in the short beam experiments. The results are presented in Figure~\ref{fig:rice_results}, which displays the decoding times before and after optimization for both architectures, alongside the percentage speedup for each code family and configuration.

\textbf{\textit{Color Codes}}. Consistent with the initial experiments, decoding times for Color Codes scaled significantly with both error rate and code distance. On the Cascade Lake architecture, the cumulative speedup after applying all optimizations ranged from 27.0\% to 37.7\%. The peak speedup of 1.61$\times$ was achieved under the \textit{SI1000} noise model at $p=0.002$ and $d=11$. On the Sapphire Rapids architecture, the cumulative speedup reached 15.0\%--42.3\%, with a peak of 1.73$\times$ recorded for the \textit{uniform} noise model at $p=0.002$ and $d=7$. These values are slightly lower than the peak speedup of 2.21$\times$ observed on the first architecture using the short beam configuration.

\textbf{\textit{Bivariate-Bicycle Codes}}. Bivariate-Bicycle Codes remained the most computationally intensive family, with decoding times exceeding 24,000 seconds for the most demanding configuration ($p=0.002, d=10$ on Cascade Lake). For Cascade Lake, the cumulative speedup reached 28.7\%--76.8\%, with a maximum gain of 4.31$\times$ for the \textit{NLR10} noise model at $p=0.002$ and $d=10$. On the Sapphire Rapids architecture, cumulative speedups ranged from 9.7\% to 64.3\%, with a peak of 2.8$\times$ for the same \textit{NLR10} configuration. These speedups are notably lower than the 5.24$\times$ peak achieved in the short beam experiments.

\textbf{\textit{Surface Codes}}. Surface Codes exhibited the lowest decoding latencies, particularly at lower error rates and smaller code distances. On the Cascade Lake architecture, cumulative speedups ranged from 20.2\% to 41.2\%, peaking at 1.70$\times$ for the configuration with $p=0.002$ and $d=11$. For Sapphire Rapids, the speedup reached 13.6\%--45.7\%, with a peak of 1.84$\times$ at $p=0.002$ and $d=9$. These results are lower than the peak speedup of 2.66$\times$ recorded on the first architecture.

\textbf{\textit{Transversal CNOT Protocols}}. Decoding times for Transversal CNOT Protocols remained high for larger error rates and distances. On the Cascade Lake architecture, cumulative speedups reached 36.9\%--46.7\%, with a peak of 1.88$\times$ at $p=0.001$ and $d=9$. On the Sapphire Rapids architecture, speedups ranged from 27.4\% to 55.3\%, reaching a peak of 2.24$\times$ at $p=0.001$ and $d=11$. As with the other families, these gains are lower than the peak of 2.75$\times$ achieved using the short beam configuration.

These results demonstrate that our strategies for accelerating the \Tesseract{} decoder are robustly portable across both hardware architectures and search configurations. Although the peak speedups on the two additional platforms did not reach the maximum values observed on the first architecture—which utilized a more restricted short beam—the cumulative gains remain highly significant. For Color Codes, Surface Codes, and Transversal CNOT Protocols, peak speedups consistently approached 2$\times$, even exceeding 2$\times$ for complex Transversal CNOT configurations. These improvements are highly practical and comparable to the 2.5$\times$ gains achieved during our initial experiments. Notably, for Bivariate-Bicycle Codes, we achieved a peak speedup of 4.31$\times$ on the Cascade Lake architecture, which closely rivals the 5.24$\times$ peak recorded in the short beam configuration. Even on the Sapphire Rapids architecture, speedups approached 3$\times$, representing a substantial reduction in computational overhead.

The success of these experiments across diverse environments is particularly noteworthy. By identifying and addressing bottlenecks on a specific architecture and configuration, and then successfully translating those gains to different CPU architectures and expanded search spaces, we have validated the fundamental nature of our optimizations. This cross-platform success proves that our solution does not merely exploit narrow hardware quirks, but rather addresses core algorithmic and memory-hierarchy inefficiencies inherent in the decoding process.

In the following section, we analyze the memory efficiency gains achieved by optimizing the critical decoding bottleneck within \Tesseract{}: the \getdetcost{} heuristic function. We detail how our optimizations targeted specific inefficiencies within this function, which guides the \astar{} search toward optimal solutions and accounts for the majority of total decoding time.

\subsection{Analyzing Cache Efficiency within the Heuristic Bottleneck}\label{sec:cache_performance}
To complement our macro-scale performance results, we conducted a fine-grained analysis of data locality and cache efficiency within the critical \getdetcost{} bottleneck. As detailed in Section~\ref{sec:optimizations}, this heuristic function was optimized through two primary strategies: \opttwo{}, which transitioned the memory layout from a Structure of Arrays (SoA) to an Array of Structures (AoS) to ensure contiguous data residency, and \optthree{}, which implemented an early-exit strategy to prune redundant iterations.

In our initial short beam experiments, \opttwo{} alone yielded a cumulative speedup of $2.75\times$ for the most computationally demanding Bivariate-Bicycle configurations. This result underscores the profound impact that data organization and memory access patterns have on decoding throughput. The disproportionate effectiveness of this optimization for Bivariate-Bicycle Codes stems from the fact that \getdetcost{} constitutes the primary bottleneck for this family, consuming 70\% to 90\% of the total decoding time. The function also represents a substantial portion of the computational load in other families, accounting for over 60\% of the decoding time in Color Codes and approximately 40\% in Surface Codes and Transversal CNOT Protocols.

To further investigate the mechanics of these improvements, we utilized the \textit{perf} tool to measure hardware performance counters for L1 and Last Level Cache (LLC) miss rates. We measured these metrics for the long beam experiments conducted on the Rice University NOTS cluster, comparing the baseline performance against the state after applying both \opttwo{} and \optthree{}. We evaluated cache miss rates across both the Cascade Lake and Sapphire Rapids microarchitectures for every tested code configuration. This micro-architectural profiling enabled us to quantify the improvements in data locality and provides a rigorous explanation for the significant speedups observed; by reducing the number of cache lines required per evaluation and minimizing expensive trips to main memory, our optimizations directly address the memory-wall challenges inherent in high-distance quantum decoding.

\subsubsection{Cache Performance for the Cascade Lake Architecture}
\begin{figure*}
\centering
\subfloat[\centering]{\includegraphics[width=.5\textwidth]{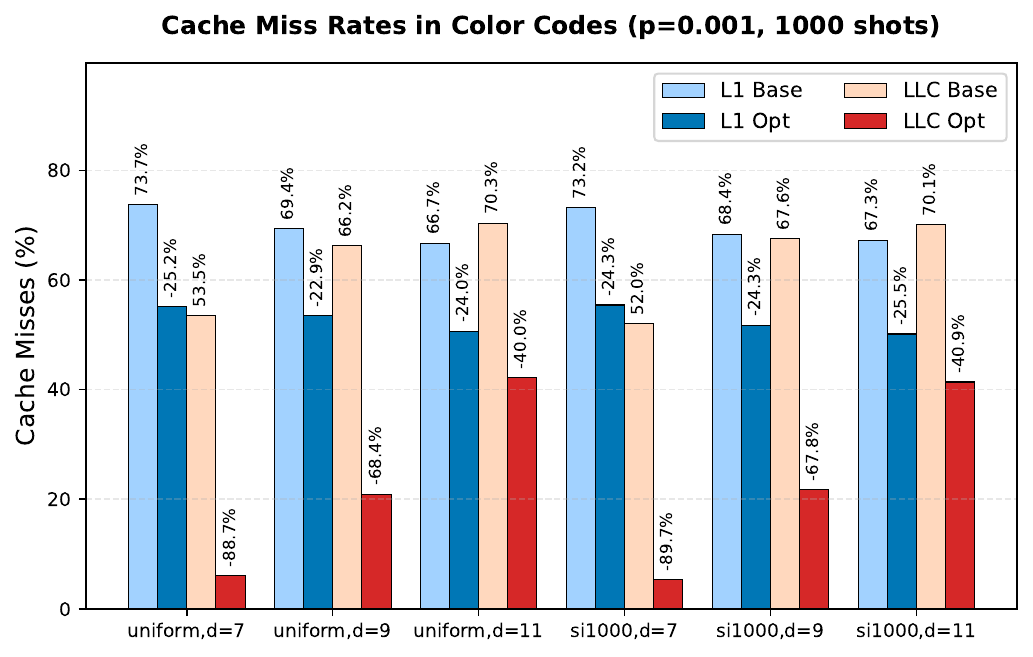}}
\subfloat[\centering]{\includegraphics[width=.5\textwidth]{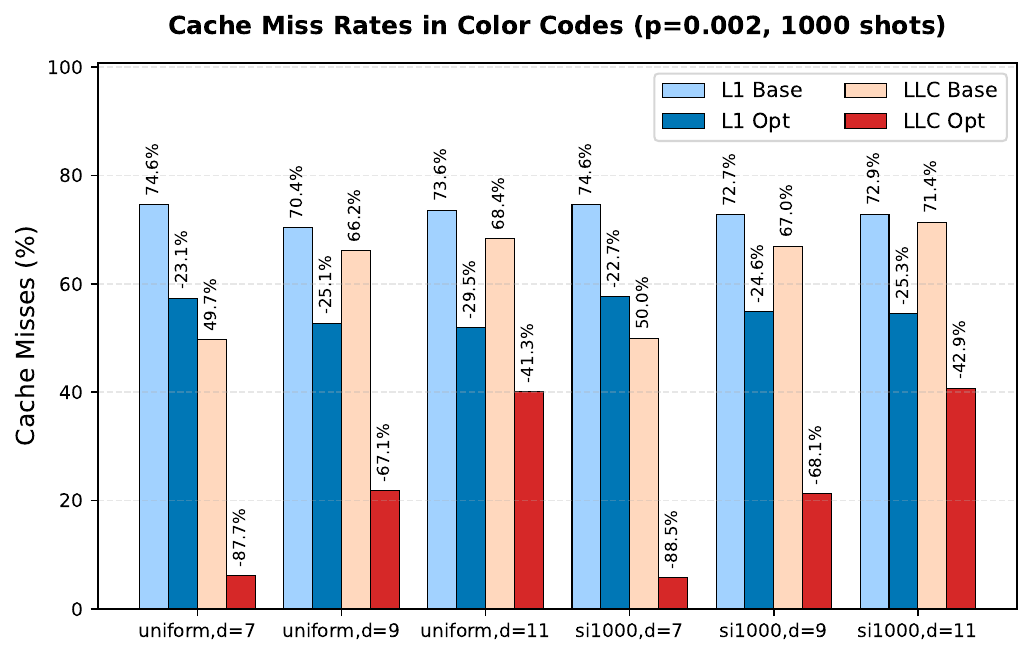}}\\
\subfloat[\centering]{\includegraphics[width=.5\textwidth]{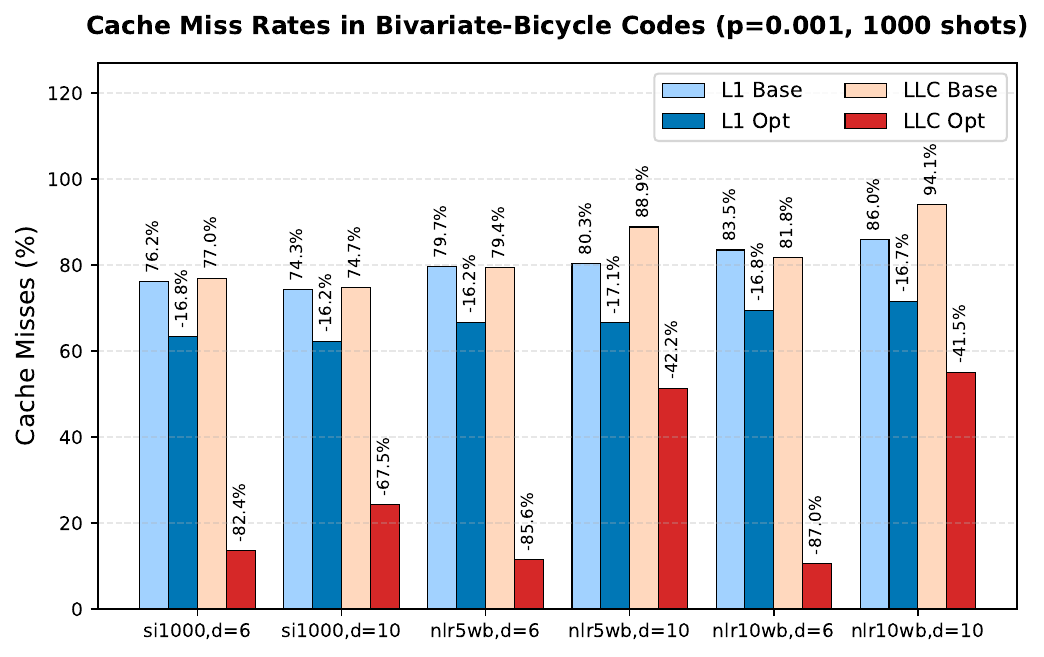}}
\subfloat[\centering]{\includegraphics[width=.5\textwidth]{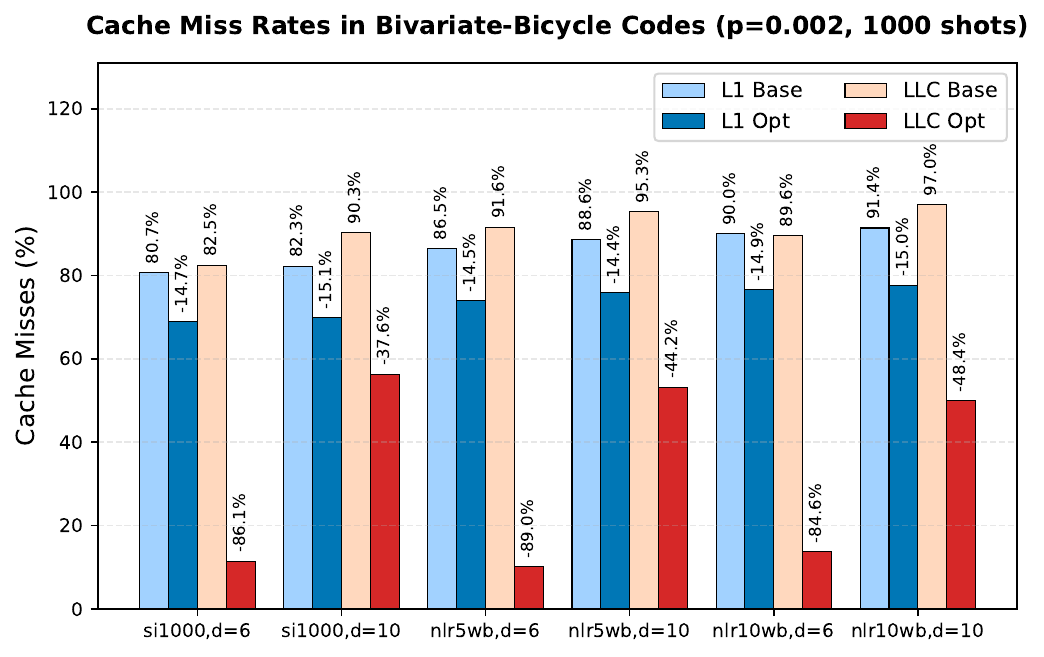}}\\
\subfloat[\centering]{\includegraphics[width=.5\textwidth]{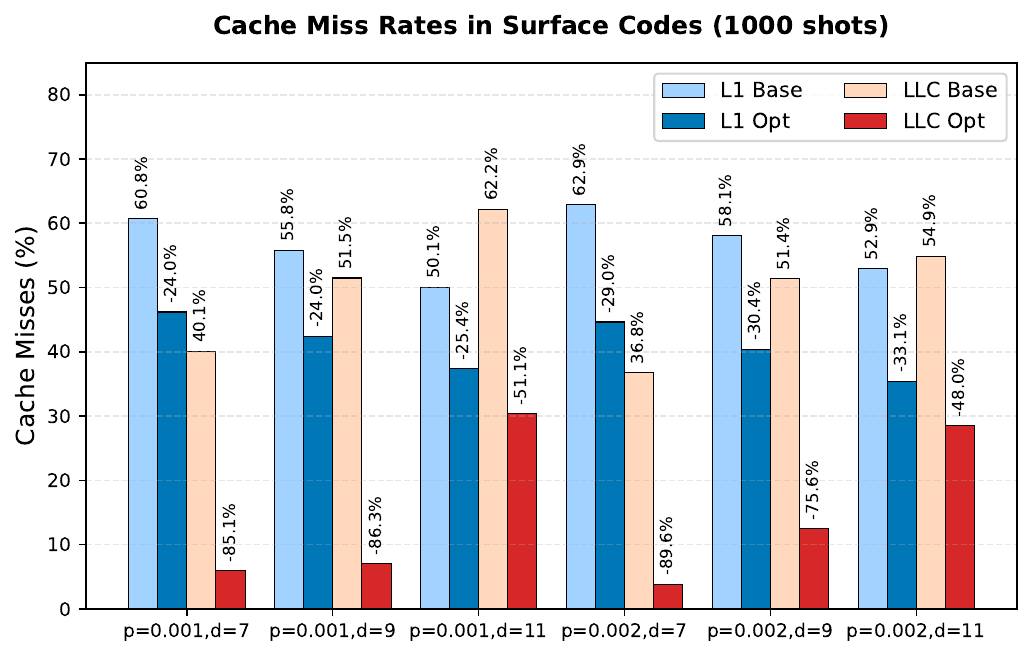}}
\subfloat[\centering]{\includegraphics[width=.5\textwidth]{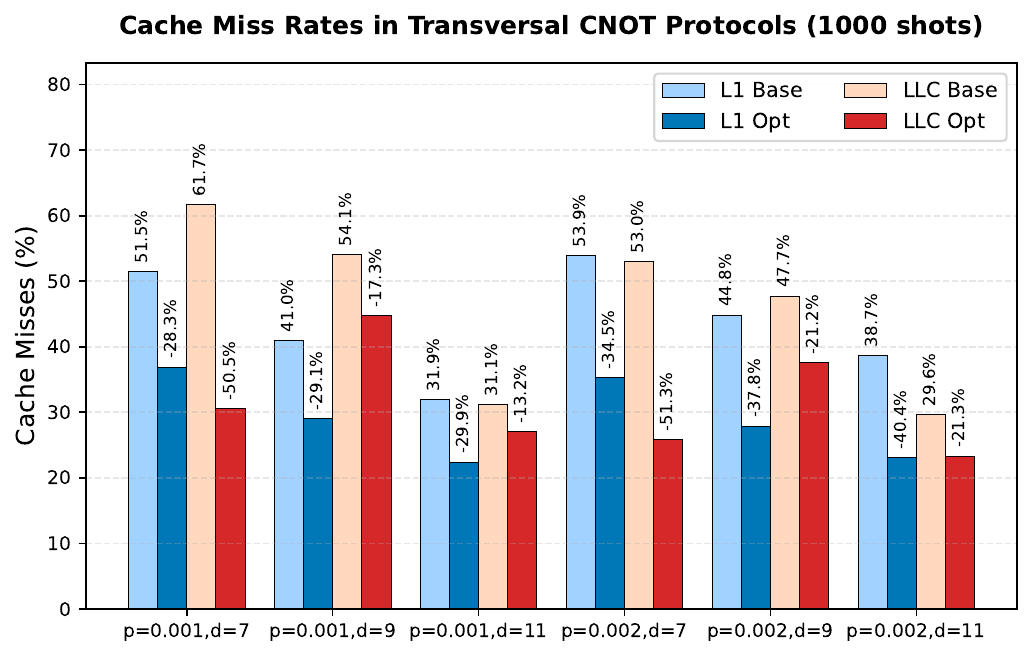}}
\caption{Cache miss rate reduction within the \getdetcost{} kernel for long beam experiments on the Cascade Lake architecture: (\textbf{a}) Color Codes at $p=0.001$; (\textbf{b}) Color Codes at $p=0.002$; (\textbf{c}) Bivariate-Bicycle Codes at $p=0.001$; (\textbf{d}) Bivariate-Bicycle Codes at $p=0.002$; (\textbf{e}) Surface Codes; (\textbf{f}) Transversal CNOT Protocols for Surface Codes.}
\label{fig:cascadelake_cache_misses}
\end{figure*}

Figure~\ref{fig:cascadelake_cache_misses} illustrates the cache miss rates for the long beam experiments on the Cascade Lake architecture. For each code family and configuration, we measured L1 and LLC misses within the \getdetcost{} kernel before and after applying \opttwo{} and \optthree{}. The results demonstrate a consistent and significant reduction in both metrics, with LLC misses decreasing by up to 90\% for configurations with lower error rates and code distances. It is remarkable that reorganizing the memory layout of data consumed by a major functional bottleneck can yield such a profound impact on data locality and system performance. In Figure~\ref{fig:cascadelake_cache_misses}, the bars represent the optimized cache miss rates, with labels indicating the percentage of improvement relative to the baseline.

\textbf{Color Codes.} Baseline L1 miss rates decreased from 66.7\%--73.7\% to 50.1\%--57.7\% after optimization, with a peak reduction of 29.5\% observed for the \textit{uniform} model at $p=0.002$ and $d=11$. LLC miss rates fell from 49.7\%--71.4\% to 5.4\%--42.2\%, representing a peak improvement of 89.7\% for the \textit{SI1000} model at $p=0.001$ and $d=7$. These results quantify the significant impact of \opttwo{} on higher-level cache residency.

\textbf{Bivariate-Bicycle Codes.} L1 miss rates, which were initially quite high (74.3\%--91.4\%), dropped to 62.3\%--77.7\% post-optimization, with a peak reduction of 17.1\% for the \textit{NLR5} model at $p=0.001$ and $d=10$. LLC miss rates plummeted from 74.7\%--97.0\% to 10.1\%--56.4\%, achieving a peak reduction of 89.0\% for the \textit{NLR5} model at $p=0.002$ and $d=6$. These findings align with our observation that \getdetcost{} is the primary bottleneck for this family.

\textbf{Surface Codes.} L1 miss rates fell from 50.1\%--62.9\% to 35.4\%--46.2\%, with a peak reduction of 33.1\% at $p=0.002$ and $d=11$. Baseline LLC miss rates dropped from 36.8\%--62.2\% to 3.8\%--30.4\%, achieving a peak reduction of 89.6\% at $p=0.002$ and $d=7$.

\textbf{Transversal CNOT Protocols.} L1 miss rates were reduced from 31.9\%--53.9\% to 22.4\%--36.9\%, with a 40.4\% peak reduction at $p=0.002, d=11$. LLC miss rates dropped from 29.6\%--61.7\% to 23.3\%--44.7\%, a peak improvement of 51.3\% at $p=0.002, d=7$. The lower baseline miss rates in this family correlate with \getdetcost{} constituting a smaller fraction of the total decoding time (around 40\% for most code configurations).

\subsubsection{Cache Performance for the Sapphire Rapids Architecture}
\begin{figure*}[t]
\centering
\subfloat[\centering]{\includegraphics[width=.5\textwidth]{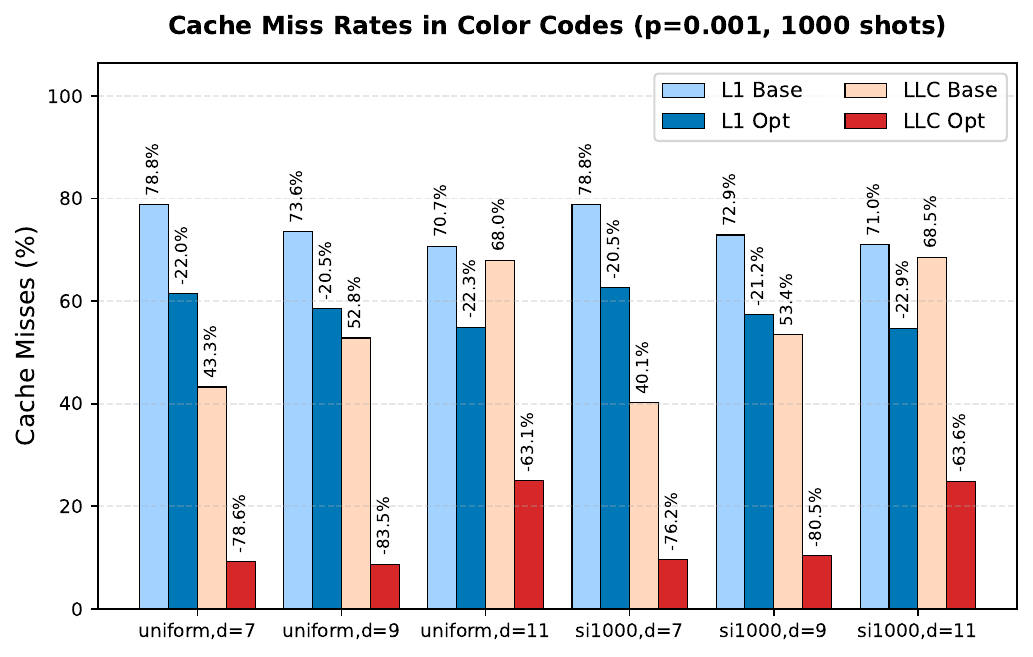}}
\subfloat[\centering]{\includegraphics[width=.5\textwidth]{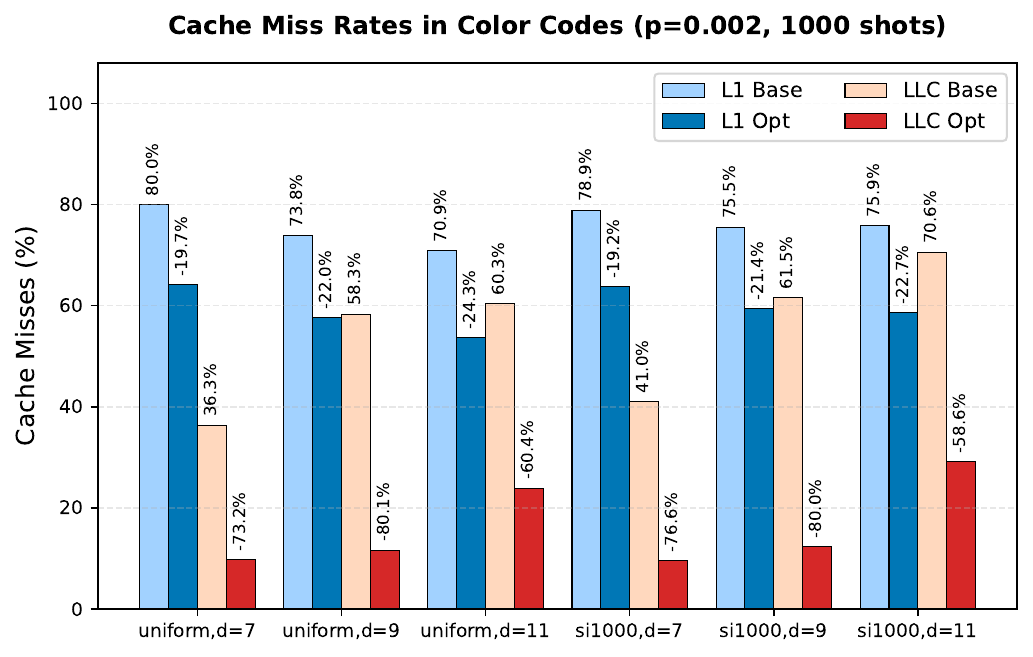}}\\
\subfloat[\centering]{\includegraphics[width=.5\textwidth]{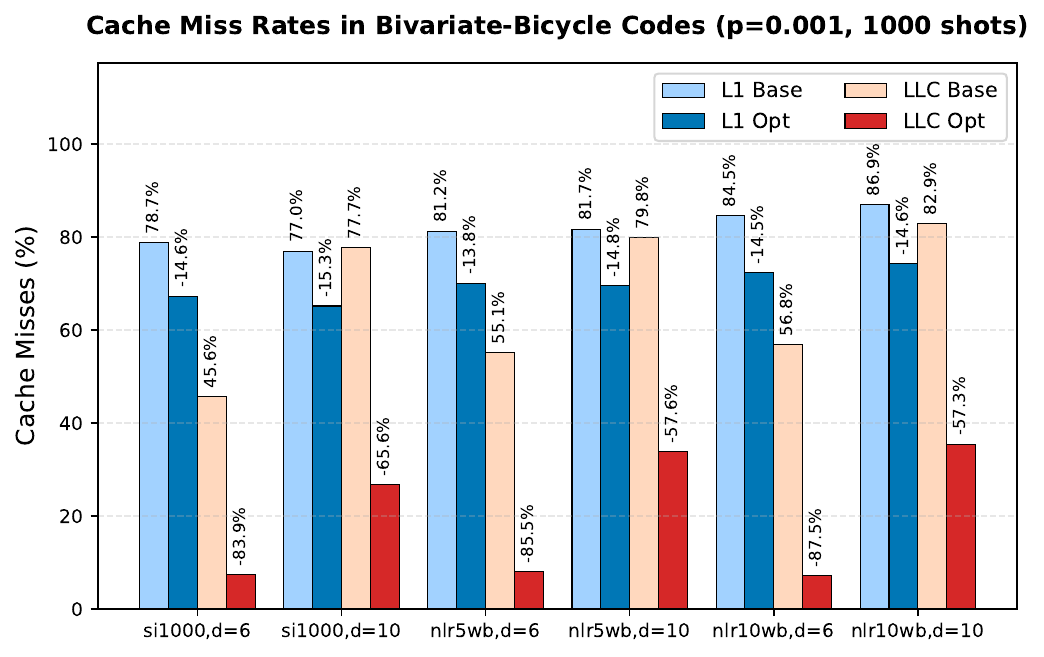}}
\subfloat[\centering]{\includegraphics[width=.5\textwidth]{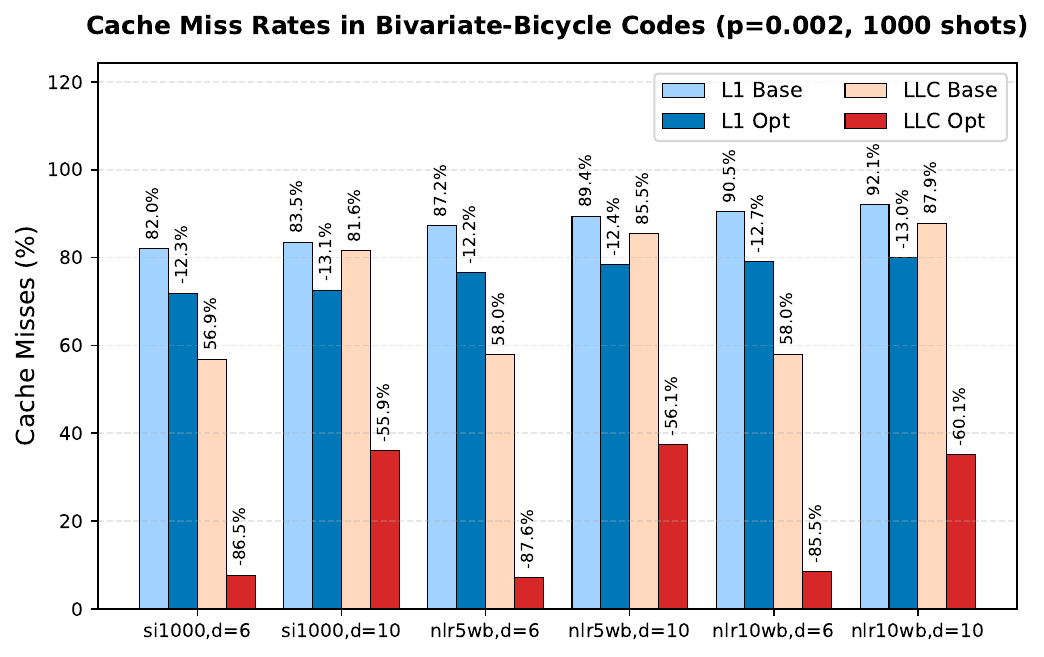}}\\
\subfloat[\centering]{\includegraphics[width=.5\textwidth]{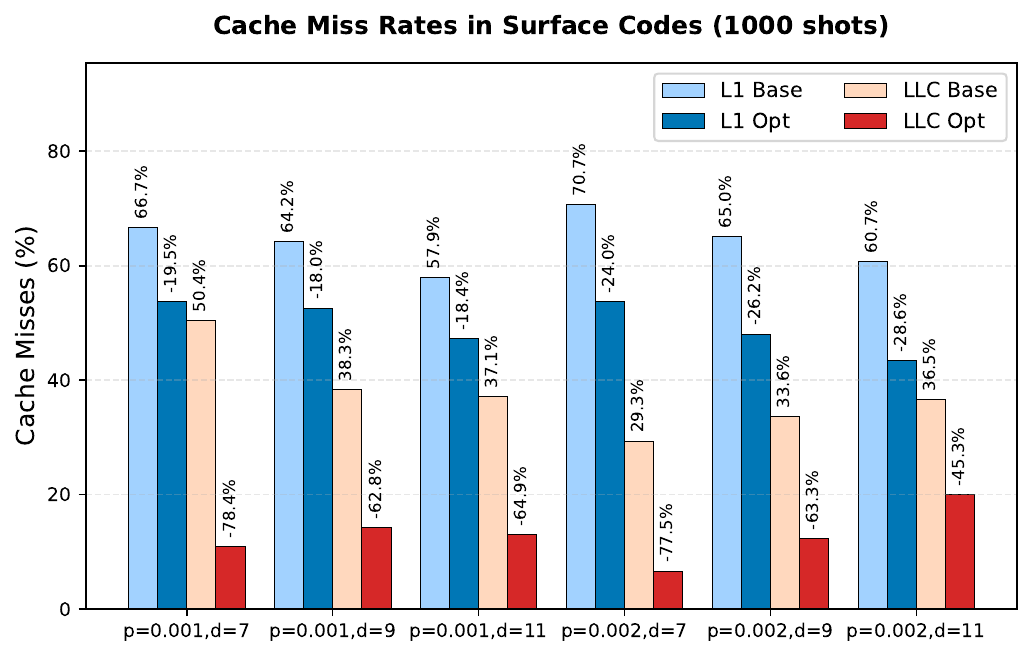}}
\subfloat[\centering]{\includegraphics[width=.5\textwidth]{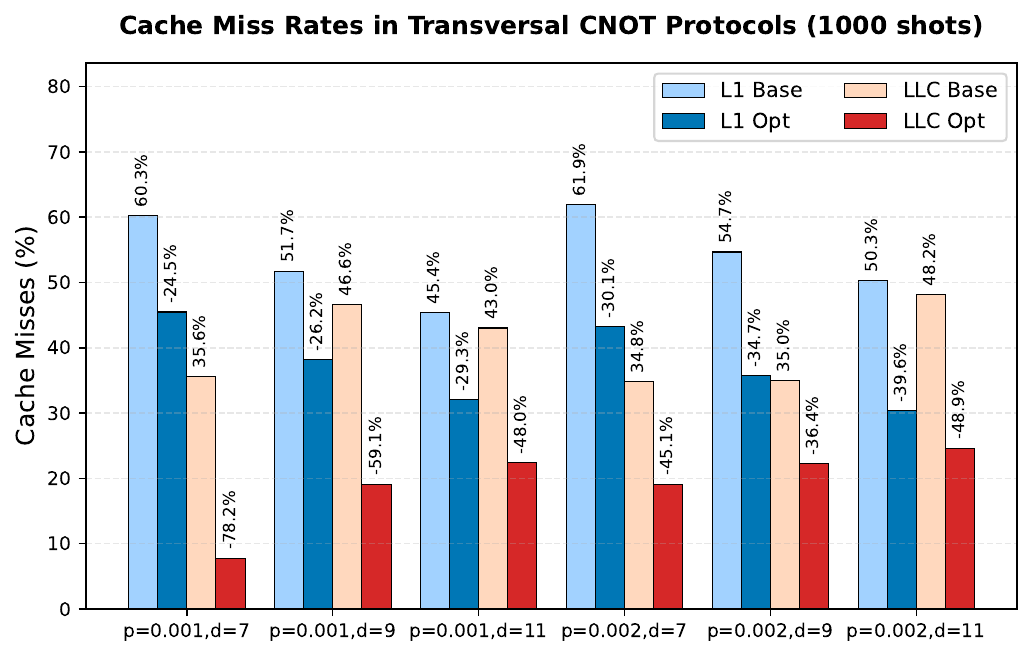}}
\caption{Cache miss rate reduction within the \getdetcost{} kernel for long beam experiments on the Sapphire Rapids architecture: (\textbf{a}) Color Codes at $p=0.001$; (\textbf{b}) Color Codes at $p=0.002$; (\textbf{c}) Bivariate-Bicycle Codes at $p=0.001$; (\textbf{d}) Bivariate-Bicycle Codes at $p=0.002$; (\textbf{e}) Surface Codes; (\textbf{f}) Transversal CNOT Protocols for Surface Codes.}
\label{fig:sapphirerapids_cache_misses}
\end{figure*}

We performed a parallel analysis for the Sapphire Rapids architecture to assess cross-platform consistency. Figure~\ref{fig:sapphirerapids_cache_misses} summarizes the cache miss rates measured within the \getdetcost{} kernel.

\textbf{Color Codes.} Baseline L1 miss rates decreased from 70.7\%--80.0\% to 53.6\%--64.2\%, with a peak reduction of 24.3\% for the \textit{uniform} model at $p=0.002$ and $d=11$. LLC miss rates fell from 36.3\%--70.6\% to 8.7\%--29.2\%, a peak reduction of 83.5\% for the \textit{uniform} model at $p=0.001$ and $d=9$.

\textbf{Bivariate-Bicycle Codes.} L1 miss rates dropped from 77.0\%--92.1\% to 65.1\%--80.0\%, a peak reduction of 15.3\% for the \textit{SI1000} model at $p=0.001$ and $d=10$. LLC misses fell significantly from 45.6\%--87.9\% to 7.1\%--37.5\%, a peak reduction of 87.6\% for the \textit{NLR5} model at $p=0.002$ and $d=6$.

\textbf{Surface Codes.} L1 miss rates fell from 57.9\%--70.7\% to 43.4\%--53.7\%, with a peak reduction of 28.6\% at $p=0.002$ and $d=11$. LLC miss rates dropped from 29.3\%--50.4\% to 6.6\%--20.0\%, with a 78.4\% peak reduction at $p=0.001$ and $d=7$.

\textbf{Transversal CNOT Protocols.} Baseline L1 miss rates fell from 45.4\%--61.9\% to 30.4\%--45.5\%, with a 39.6\% peak reduction at $p=0.002$ and $d=11$. LLC miss rates decreased from 34.8\%--48.2\% to 7.8\%--24.6\%, with a 78.2\% peak reduction at $p=0.001$ and $d=7$.

The dramatic reduction in LLC miss rates—reaching up to 90\%—stems from the synergy between enhanced spatial locality and the CPU's hardware prefetching mechanisms. By transitioning to an AoS layout, we ensured that the data fields required for a single \getdetcost{} evaluation are packed into contiguous memory blocks. This alignment enables a single cache line fetch to populate the L1 cache with all necessary parameters, effectively eliminating the fragmented, non-contiguous memory requests that previously bottlenecked the LLC. Furthermore, by aligning our data structures with 64-byte cache line boundaries, we maximized the efficiency of the hardware prefetcher; this enables the CPU to accurately predict and load subsequent data blocks into the cache hierarchy before they are explicitly requested.

\section{Conclusion}
The implementation of robust and efficient decoders is a fundamental requirement for the success of practical quantum error correction. In this paper, we presented a systematic investigation into the performance of the \Tesseract{} decoder, a search-based decoder that utilizes \astar{} search to navigate the exponentially large space of error configurations. While \Tesseract{} provides high accuracy and significant speed advantages over traditional Integer Programming decoders, our work demonstrates that its performance can be further elevated through rigorous profiling and low-level optimizations.

By conducting a detailed bottleneck analysis, we identified and resolved critical inefficiencies within the decoder's \Cpp{} implementation. We iteratively applied four optimization strategies, focusing on improving memory access patterns, leveraging advanced data structures, and accelerating the \getdetcost{} heuristic kernel by reorganizing memory layouts from a Structure of Arrays (SoA) to an Array of Structures (AoS). Our results show that such systematic refinements to data locality and cache efficiency yield profound performance gains. Across three distinct hardware architectures, we achieved consistent speedups of approximately 2$\times$ for Color codes, Surface codes, and Transversal CNOT protocols, often exceeding 2.5$\times$. For the most computationally demanding configurations of Bivariate-Bicycle codes, we achieved a peak speedup exceeding 5$\times$ on the primary architecture.

Ultimately, these enhancements make \Tesseract{} a more scalable and viable tool for decoding the large code distances and high error rates required for fault-tolerant quantum computing. Our work serves as a practical case study in high-performance engineering for \QEC{}, providing a foundation and an extensive performance dataset to guide the development of future real-time decoding solutions.

\section*{Author Contributions}
Conceptualization, D.G., L.A.B., and N.S.; methodology, D.G. and N.S.; software, D.G.; validation, D.G., L.A.B., and N.S.; formal analysis, D.G.; investigation, D.G.; resources, L.A.B. and N.S.; data curation, D.G.; writing---original draft, D.G.; writing---review and editing, D.G., L.A.B., and N.S.; visualization, D.G.; supervision, L.A.B. and N.S.; project administration, L.A.B.; funding acquisition, L.A.B.

\section*{Data Availability}
The \Tesseract{} decoder is an open-source project available at~\cite{tesseract-repo}. The specific implementations of the optimizations, including detailed descriptions and performance benchmarking results, are publicly archived in the repository's pull requests:~\cite{opt1},~\cite{opt2_3}, and~\cite{opt4}. All other data supporting the findings of this study are included within the article.

\section*{Acknowledgments}
The majority of this work was performed at Google Quantum AI during D.G.’s internship under the supervision of L.A.B. and N.S. The authors thank the Google Quantum AI team for providing the internship opportunity and the computational resources used for the primary phase of the study, which included the initial decoder profiling, optimization design, and experiments on the first hardware architecture. Additional evaluation experiments on the second and third architectures were conducted at Rice University; this work was supported in part by the NOTS cluster operated by Rice University's Center for Research Computing (CRC).

\bibliographystyle{plainnat}
\bibliography{references}

\end{document}